\documentclass[twoside,12pt]{report}
\usepackage{graphics}
\usepackage{epsfig,subfigure}
\usepackage[centertags]{amsmath}
\usepackage{amsfonts}
\usepackage{amssymb}
\usepackage{amsthm}
\usepackage{newlfont}
\usepackage{XThesis} 
\usepackage{XTocinc} 
\newlength{\defbaselineskip}
\newcommand{\setlinespacing}[1]%
           {\setlength{\baselineskip}{#1 \defbaselineskip}}

\newcommand{\h}{{\cal H}}

\newcommand{\dN}{\mathbb N}

\newcommand{\dR}{\mathbb R}

\newcommand{\id}{\mathbb I}

\newcommand{\dZ}{\mathbb Z}

\numberwithin{equation}{section}

\begin{document}



\dedicate{To someone without whom this Thesis\\ would be never finished}

\nolistoftables \nolistoffigures 
 \phd



\title{Examination of the nature of\\ the Bianchi type cosmological singularities.
}

\author{Piotr Dzier\.{z}ak}




{
\typeout{:?0000} 
\beforepreface
\typeout{:?1111} 
}
{
\typeout{Acknowledgements}

\prefacesection{Acknowledgements}
\def\baselinestretch{1.0}
\setlinespacing{1.15}

I would like to thank Prof. W\l odzimierz Piechocki, my
supervisor, who introduced me to the problems and methods of
Quantum Cosmology and with whom I shared a pleasure of joint
scientific investigations.

\noindent I am grateful to my dear friend Dr Przemyslaw Ma\l kiewicz, who helped me understanding some intriguing mathematical problems.\\

\noindent
Warsaw, December 30, 2010 \hfill Piotr Dzier\.{z}ak\\

\bigskip

\goodbreak

\bigskip

\goodbreak

\noindent This Thesis is based on the following author's papers:\\

\noindent [1] P.Dz., Jacek Jezierski, Przemys\l aw Ma\l kiewicz and W\l odzimierz Piechocki,\\\indent ``The minimum length problem of loop quantum cosmology'', Acta Phys. Polon. \indent \textbf{41} (2010) 717, arXiv:0810.3172.

\noindent [2] P.Dz., Przemys\l aw Ma\l kiewicz and W\l odzimierz Piechocki, \\\indent``Turning Big Bang into Big Bounce: I. Classical Dynamics'', Phys.\ Rev.\  D {\bf 80}, \indent 104001 (2009), arXiv:0907.3436.

\noindent [3] P.Dz. and W\l odzimierz Piechocki, \\\indent ``Bianchi I model in terms of non-standard LQC: Classical dynamics'', Phys.\ Rev.\  \indent D {\bf 80}, 124033 (2009), arXiv:0909.4211.

\noindent [4] P.Dz. and W\l odzimierz Piechocki, \\\indent ``Bianchi I model of the universe in terms of nonstandard LQC'', Ann. Phys. \textbf{19} \indent (2010) 290.

\noindent [5] Przemys\l aw Ma\l kiewicz, W\l odzimierz Piechocki and P.Dz., \\\indent``Bianchi I model in terms of nonstandard loop quantum cosmology: Quantum \indent dynamics.'', Class.\ Quant.\ Grav. {\bf 28}, 085020 (2011), arXiv:1010.2930.

}
\afterpreface
\def\baselinestretch{1}
\setlinespacing{1.66}
{ \typeout{Introduction}

\nonumchapter{Introduction}

It results from cosmological observations that the Universe emerged from a state characterized by extremely high density of matter fields called Big Bang or cosmological singularity \cite{MTW,SWH,JPAK,JMMS}. The observational data are in comparatively good agreement with calculations obtained within the standard cosmological models (SCM). The latter is based on General Relativity (GR), and includes dark matter, dark energy and the inflation transition. However, SCM provides a phenomenological description. Deep understanding of the cosmological data, in particular of an early Universe requires an elementary quantum theory being able to explain creation of spacetime filled with matter fields.

Probably, the best starting point for finding such a theory is quantization of the BKL (Belinskii, Khalatnikov, Lifshitz) scenario \cite{BKL1,BKL2,BKL}. First of all, it is clear from this scenario that a general solution of the Einstein equation with the cosmological singularity does exist. By this term we mean a singularity in time on spacelike hypersurface characterized by blowing up the curvature invariants together with diverging energy densities of matter fields. The BKL solution of GR is general and stable. By general we mean containing a non-zero measure subset of all initial conditions. Stability means that no infinitesimal perturbations of initial conditions is able to change the singular character of the solution.

Secondly the BKL scenario says that in asymptotic vicinity of the singularity this general solution has complicated oscillatory behavior of chaotic character. There is considerable support for this scenario both from analytical \cite{PrzegBKL1, PrzegBKL2} and numerical \cite{Numeryczna1} investigations. What is important in the BKL the dynamics at any spatial point can be approximated by that of the homogeneous (but in general non-isotropic) models which are called Bianchi models \cite{BKL1, BKL2, BKL}.

In the BKL scenario, time derivatives of gravitational field is shown to dominate over spatial derivatives for long stretches of time. During such periods called Kasner epochs, an evolution of gravitational field may be approximated by the Bianchi I model \cite{PrzegBKL3}.

In each Kasner epoch the Universe is approximated by the Bianchi I metric with some specific set of parameters $\textrm{k}_i$ satisfied equations $\sum_{i=1}^{3} \textrm{k}_i= 1$ and $\sum_{i=1}^{3} \textrm{k}_i^2 + \textrm{k}_{\phi}^2=1$ where $\textrm{k}_{\phi}$ describes density of matter fields. There exist two classes of solutions. The first, called Kasner-like, one of $\textrm{k}_i$ has different sign than two others. It means that the Universe contracts in two directions and expands in the third. The second one, called Kasner-unlike, all the parameteres have the same sign, which means that there is a contraction in every direction.
This is not the end, because from time to time there occur short periods in the evolution when spatial derivatives of gravitational field dominate over time derivatives, which lead to the transitions between Kasner epochs. In each transition the set of the parameters $\textrm{k}_i$ characterizing the specific Bianchi I model changes to another one. Dynamics of the transitions can be modelled by the Bianchi II time evolution \cite{BianchiII}.

In the BKL scheme, the Universe may undergo an infinite number of chaotic-like transitions (depending on equation of state of matter fields) from one Kasner epoch to another, and finally collapse to a singularity in a finite proper time \cite{PrzegBKL1, PrzegBKL2}. It is clear that before the system approaches the singularity, the spacetime curvature acquires the Planck scale so the classical scenario cannot be trusted.

It is clear that, according to the BKL scenario, there are two basic steps in understanding of an early Universe. The first one means the construction of the quantum Bianchi I model of the Universe \cite{Ashtekar:2009vc}, and the second one means finding the quantum Bianchi II model.
This Thesis is devoted to the first task.

Quantization in this Thesis is based on the so-called nonstandard Loop Quantum Cosmology (LQC) which is an alternative to the standard Loop Quantum Cosmology. The difference between them will be discussed later. The standard LQC is a cosmological counterpart of Loop Quantum Gravity (LQG) and is is obtained by symmetry reduction of LQG, which as a field theory has infinite number of degrees of freedom \cite{Ashtekar:2003hd}. On the other hand LQG is a non-perturbative theory being today one of the most promising candidate to the theory of unification GR with quantum physics. Both geometry and matter are dynamical and described by quantum mechanics. In LQG there is no background spacetime \cite{Ashtekar:2004eh}.

This Thesis is organized as follows:\\
Chapter 1 is an introduction and contains main features of both existing LQC methods, namely the standard and the nonstandard. In Section 1 of this chapter we describe formalism of the standard LQC concentrating mainly on its successes in resolving the cosmological singularity. Section 2 contains a motivation for another LQC method, which underlines the Thesis. In Section 3 we describe this method, called the nonstandard LQC and make a comparison between these two methods. In Section 4 we present an application of the nonstandard LQC to the simplest cosmological model, namely FRW. At the classical level we show the occurrence of Big Bounce. At the quantum level particular attention is paid to volume and energy density operators. Spectra of these operators are analyzed in details.\\
Chapter 2 is the main part of the Thesis. Here we present an application of nonstandard LQC to the Bianchi I cosmological model. Section 1 is devoted to calculations done on the classical level, which mean solving the equations of motions and finding the algebra of elementary observables. In Section 2 we make, before quantizing, some comfortable redefinitions and, what is more important, analyze structure of the phase space. So-called true Hamiltonian is introduced. In Section 3 we face the quantization in Kasner-like and Kasner-unlike cases. Particular attention is paid to the volume operator and the problem of an evolution.\\
In Conclusions we make summary of all the results.\\
Appendix A presents curvature of connection expressed in terms of holonomies. Classification of phase space in terms of observables includes Appendix B. Some ambiguities in quantization are disscussed in Appendix C.

\setlinespacing{1.66}

\def\baselinestretch{1}

\chapter{Loop Quantum Cosmology}

\def\baselinestretch{1.66}


\smallskip

\goodbreak
\section{Formalism of standard LQC}

\subsection{Basic facts}

By the standard LQC one means LQC that is strongly inspired by LQG \cite{Ashtekar:2004eh, TT, CR}. The inspiration consists mainly in applying the two ingredients of LQG: (i) modification of the curvature of connection by loop geometry, and (ii) making use of the holonomy-flux variables. The construction of LQC has been carried out by mimicry of the construction of LQG.

The LQC was firstly applied to quantization of FRW model as FRW is the most common model in cosmology and astrophysics today. This field is not very old, because the first papers appeared less than 10 years ago \cite{Bojowald:2001xe,Bojowald:2003md}. Since then more than 100 papers have been written. Amongst the most important are \cite{Ashtekar:2003hd,Bojowald:2006da,Ashtekar:2006uz,Ashtekar:2006wn,Ashtekar:2007em}. We recommend also a few review articles \cite{Przeg1,Przeg2,Przeg3,Ashtekar:2008vv}.

The standard LQC \cite{Ashtekar:2003hd,Bojowald:2006da} means basically the Dirac method of quantization, which begins with quantization of the kinematical phase space followed by
imposition of constraints of the gravitational system in the form of operators acting on a kinematical Hilbert space. Then, one must solve these constraints which means finding kernels of the operators. The kernels are used to identify the physical Hilbert space.

\subsection{Big Bounce}

The most important result obtained within the standard LQC is that the classical Big Bang (and also the Big Crunch) is replaced by the quantum Big Bounce due to strong quantum effects at the Planck scale \cite{Przeg1,Przeg2,Przeg3}. The result was obtained for FRW models with $k=0$ and $k=1$ and for Bianchi I model \cite{Przeg1}. It was done using analytical and numerical methods \cite{Bojowald:2001xe,Ashtekar:2006wn,Date:2004fj}. In all these models the role of internal time was played by a scalar field which enables interpreting the Hamiltonian constraint as an evolution equation. Singularity is resolved in the sense that observables like energy density of matter which classically diverge, are represented by operators bounded from above on the states (vectors of the physical Hilbert space) which are semi-classical asymptotically \cite{Ashtekar:2006wn}. It is suggested in \cite{Ashtekar:2007em} that the bounce may occur for the states which are more general than
semi-classical at late times, which demonstrates robustness of LQC results. Quantum evolution, described by (\ref{motquanD}), is deterministic across the bounce region. An universe undergoes a bounce during the evolution from an epoch before the Big Bang to an epoch after the Big Bang, so on the ``other side'' of the bounce there is also a universe \cite{Przeg2}. These are main highlights of LQC (see, e.g. \cite{Ashtekar:2008vv} for a complete list).

There exists an alternative to the standard LQC approach called the nonstandard LQC, which is presented and applied in the rest of the Thesis.

\section{Motivation for nonstandard LQC}

\subsection{Hamiltonian}

The gravitational part of the classical Hamiltonian, $H_g$, in GR is a linear combination of the first-class constraints, and reads
\cite{Ashtekar:2003hd,Ashtekar:2004eh,TT,CR,Bojowald:2006da}
\begin{equation}\label{motham1}
    H_g:= \int_\Sigma d^3 x (N^i C_i + N^a C_a + N C),
\end{equation}
where $\Sigma$ is the spacelike part of spacetime $\dR \times
\Sigma$, $~(N^i, N^a, N)$ denote Lagrange multipliers, $(C_i, C_a,
C)$ are the Gauss, diffeomorphism and scalar constraints.
In our notation  $(a,b = 1,2,3)$ are spatial, and $(i,j,k = 1,2,3)$
internal $SU(2)$ indices. The constraints must satisfy a specific
algebra.

For flat FRW model with massless scalar field we can rewrite the gravitational part of the classical Hamiltonian, having fixed local gauge and diffeomorphism freedom, in the form (see, e.g. \cite{Ashtekar:2006wn})
\begin{equation}\label{mothamG}
H_g = - \gamma^{-2} \int_{\mathcal V} d^3 x ~N
e^{-1}\varepsilon_{ijk}
 E^{aj}E^{bk} F^i_{ab}\, ,
\end{equation}
where  $\gamma$ is the Barbero-Immirzi parameter, $\mathcal
V\subset \Sigma$ is an elementary cell, $\Sigma$ is spacelike
hyper-surface, $N$ denotes the lapse function,
$\varepsilon_{ijk}$ is the alternating tensor, $E^a_i $ is a
densitized  vector field, $e:=\sqrt{|\det E|}$, and where
$F^i_{ab}$ is the curvature of an $SU(2)$ connection $A^i_a$.

The resolution of the singularity, obtained within LQC, is based
on rewriting the curvature $F^k_{ab}$ in terms of holonomies
around loops. The curvature $F^k_{ab}$ may be determined
\cite{Ashtekar:2006wn} by making use of the formula (see the Appendix A)
\begin{equation}\label{motcur}
F^k_{ab}= -2~\lim_{Ar\,\Box_{ij}\,\rightarrow \,0}
Tr\;\Big(\frac{h^{(\lambda)}_{\Box_{ij}}-1}{\lambda^2
V_o^{2/3}}\Big)\;{\tau^k}\; ^o\omega^i_a  \; ^o\omega^j_a ,
\end{equation}
where
\begin{equation}\label{motbox}
h^{(\lambda)}_{\Box_{ij}} = h^{(\lambda)}_i h^{(\lambda)}_j (h^{(\lambda)}_i)^{-1}
(h^{(\lambda)}_j)^{-1}
\end{equation}
is the holonomy of the gravitational connection around the square
loop $\Box_{ij}$,  considered over a face of the elementary cell,
each of whose sides has length $\lambda V_o^{1/3}$
with respect to the flat fiducial metric $^o q_{ab}:=
\delta_{ij}\, ^o \omega^i_a\, ^o \omega^j_a $; fiducial triad $^o
e^a_k$ and cotriad $^o \omega^k_a$ satisfy $^o \omega^i_a\,^o
e^a_j = \delta^i_j$; the spatial part of the FRW metric is
$q_{ab}=a^2(t)\,^o q_{ab}$; $~Ar\,\Box_{ij}$ denotes the area of
the square; $V_o = \int_{\mathcal V} \sqrt{^o q} d^3 x$ is the
fiducial volume of $\mathcal V$. Because its value is for our analysis not essential, we set $V_0=1$. Here we would like to emphasize that (\ref{motcur}) is exact iff $\lambda=0$, which is not our case.

The holonomy along straight edge $ ^oe^a_k \partial_a $ of length
$\lambda V_0^{1/3}$ reads
\begin{equation}\label{mothol}
h^{(\lambda)}_k (c) = \mathcal{P} \exp \,\big(\int_0^{\lambda V_0^{1/3}}
\tau_{(k)} A^{(k)}_a dx^a \big) = \exp (\tau_{k}\lambda c) =\cos (\lambda
c/2)\;\id + 2\,\sin (\lambda c/2)\;\tau_k,
\end{equation}
where $\tau_k = -i \sigma_k/2\;$ ($\sigma_k$ are the Pauli spin
matrices) and  $\mathcal{P}$ denotes the path ordering symbol.
Equation (\ref{mothol}) presents the holonomy calculated in the
fundamental, $j=1/2$, representation of $SU(2)$.

Making  use of (\ref{mothamG}), (\ref{motcur})  and the so-called
Thiemann identity \cite{TT}
\begin{equation}\label{motidentT}
\varepsilon_{ijk}\,e^{-1}\,E^{aj}E^{bk} = \frac{sgn(p)}{2\pi G
\gamma \lambda}\,\sum_k\,^o\varepsilon^{abc}\,
^o\omega^k_c\,Tr \Big(h_k^{(\lambda)}\{(h_k^{(\lambda)})^{-1},V\}\,\tau_i
\Big)
\end{equation}
leads to $H_g$ in the form
\begin{equation}\label{mothamR}
    H_g = \lim_{\lambda\rightarrow \,0}\; H^{(\lambda)}_g ,
\end{equation}
where
\begin{equation}\label{mothamL}
H^{(\lambda)}_g = - \frac{sgn(p)}{2\pi G \gamma^3 \lambda^3}
\sum_{ijk}\,N\, \varepsilon^{ijk}\, Tr \Big(h^{(\lambda)}_i
h^{(\lambda)}_j (h^{(\lambda)}_i)^{-1} (h^{(\lambda)}_j)^{-1}
h_k^{(\lambda)}\{(h_k^{(\lambda)})^{-1},V\}\Big),
\end{equation}
and where $V= |p|^{\frac{3}{2}}= a^3$ is the volume of the
elementary cell $\mathcal{V}$.

The connection $A^k_a$ and the density weighted triad $E^a_k$ which occurs in (\ref{motidentT}) is determined by the conjugate variables $c$ and
$p$ as follows: $A^k_a = \,^o\omega^k_a\,c$ and
$\,E^a_k = \,^oe^a_k\,\sqrt{q_o}\,p$, where $\,c =
\gamma \,\dot{a}$ and $\,|p| = a^2$.

It should be noticed that in this section we use the ``old'' quantization scheme \cite{Ashtekar:2006uz}. There exists also the ``improved'' scheme $\bar\mu= \sqrt{\frac{1}{|p|}}\,\lambda$ described in \cite{Ashtekar:2006wn} and used in section ``Application'' of this Thesis. It has serious advantages and is now commonly used by LQC community. However, obtained results concern both methods \cite{Dzierzak:2008dy}.

The classical total Hamiltonian for FRW universe with a massless
scalar field, $\phi$, reads
\begin{equation}\label{motham}
   H = H_g + H_\phi \approx 0,
\end{equation}
where $H_g$ is defined by (\ref{mothamR}) and where sign ``$\approx$'' reminds that $H$ is a constraint of the system. The Hamiltonian of the
scalar field  is known to be: $H_\phi = N\,p^2_\phi
|p|^{-\frac{3}{2}}/2$, where $\phi$ and $p_\phi$ are the
elementary variables satisfying $\{\phi,p_\phi\} = 1$. The
relation $H \approx 0$ defines the physical phase space of
considered gravitational system with constraints.

\subsection{Quantization}

In the Dirac quantization \cite{PAM,HT}  we find  a kernel of the
quantum operator $\hat{H}$ corresponding to $H$, i.e.
\begin{equation}\label{motker}
    \hat{H}\Psi = 0 ,
\end{equation}
since the classical Hamiltonian is a constraint of the system,
and try to define a scalar product on the space of solutions to
(\ref{motker}). This gives a  starting point for the determination of
the physical Hilbert space $\mathcal{H}_{phys}$.

\subsubsection{Kinematics}

 The classical elementary functions satisfy the relation
\begin{equation}\label{motrelA}
\{p, N_\lambda \} = -i \frac{ 4\pi G  \gamma}{3} \lambda
N_\lambda,
\end{equation}
where $G$ is the Newton constant. Quantization of the algebra
(\ref{motrelA}) is done by making use of the prescription
\begin{equation}\label{motpres}
    \{\cdot,\cdot\} \longrightarrow \frac{1}{i
    \hbar}\,[\cdot,\cdot].
\end{equation}
The basis of the representation space is chosen to be the set of
eigenvectors of the momentum operator \cite{Ashtekar:2003hd} and
is defined by
\begin{equation}\label{motrelB}
\hat{p}\,|\mu\rangle = \frac{4\pi\gamma l_p^2}{3}\, \mu
\,|\mu\rangle,~~~~\mu \in \dR ,
\end{equation}
where $l_p^2 = G \hbar$. The operator corresponding to $N_\lambda$
acts as follows
\begin{equation}\label{motrelC}
\hat{N}_\lambda  \,|\mu\rangle = |\mu + \lambda \rangle.
\end{equation}
The quantum algebra corresponding to (\ref{motrelA}) reads
\begin{equation}\label{motrelD}
    \frac{1}{i \hbar}[\hat{p},\hat{N}_\lambda ]\,|\mu\rangle = -i \frac{4\pi G
    \gamma}{3}\,\lambda\,\hat{N}_\lambda\, |\mu\rangle.
\end{equation}
The carrier space, $\mathcal{F}_g$, of the representation
(\ref{motrelD}) is the space spanned by $\{|\mu\rangle,\,\mu\in
\dR\}$ with the scalar product defined as
\begin{equation}\label{motscal}
\langle \mu |\mu ^\prime \rangle:=  \delta_{\mu,\mu^\prime},
\end{equation}
where $\delta_{\mu,\mu^\prime}$ denotes the Kronecker delta.

The completion of $\mathcal{F}_g$ in the norm induced by
(\ref{motscal}) defines the Hilbert space $\mathcal{H}^g_{kin}= L^2
(\dR_{Bohr}, d\mu_{Bohr})$, where $\dR_{Bohr}$ is the Bohr
compactification of the real line and $d\mu_{Bohr}$ denotes the
Haar measure on it \cite{Ashtekar:2003hd}. $\mathcal{H}^g_{kin}$
is  the kinematical space of the gravitational degrees of freedom.
The kinematical Hilbert space of the scalar field is
$\mathcal{H}^\phi_{kin} = L^2(\dR, d\phi)$, and the operators
corresponding to the elementary variables are
\begin{equation}\label{motelem}
(\hat{\phi}\psi)(\phi)= \phi \psi(\phi),~~~~\hat{p}_\phi \psi = -i
\hbar \frac{d}{d\phi}\psi .
\end{equation}
The kinematical Hilbert space of the gravitational field coupled
to the scalar field is  defined to be $\mathcal{H}_{kin}=
\mathcal{H}^g_{kin}\otimes \mathcal{H}^\phi_{kin}$.

\subsubsection{Dynamics}

The  resolution of the singularity
\cite{Ashtekar:2003hd,Bojowald:2006da,Ashtekar:2006uz,Ashtekar:2006wn,Ashtekar:2007em} is mainly due to the  peculiar way of defining the  quantum operator corresponding to $H_g$. Let us consider this issue in more details.

Using the prescription $\{\cdot,\cdot\}\rightarrow \frac{1}{i
\hbar} [\cdot,\cdot]$ and specific factor ordering of operators,
one obtains from (\ref{mothamL}) a quantum operator corresponding to
$H_g^{(\lambda)}$ in the form \cite{Ashtekar:2003hd}
\begin{equation}\label{mothamRQ}
\hat{H}^{(\lambda)}_g =  \frac{i\, sgn(p)}{2\pi l_p^2 \gamma^3
\lambda^3} \sum_{ijk} \varepsilon^{ijk}\, Tr
\Big(\hat{h}^{(\lambda)}_i \hat{h}^{(\lambda)}_j
(\hat{h}^{(\lambda)}_i)^{-1} (\hat{h}^{(\lambda)}_j)^{-1}
\hat{h}_k^{(\lambda)}\{(\hat{h}_k^{(\lambda)})^{-1},\hat{V}\}\Big).
\end{equation}
One can show \cite{Ashtekar:2003hd} that (\ref{mothamRQ}) can be rewritten as
\begin{equation}\label{mothamK}
\hat{H}^{(\lambda)}_g |\mu\rangle = \frac{3}{8 \pi \gamma^3
\lambda^3 l_p^2}\Big(V_{\mu + \lambda}- V_{\mu - \lambda}
\Big)\big(|\mu + 4 \lambda\rangle - 2 |\mu\rangle + |\mu - 4
\lambda\rangle\big),
\end{equation}
where $|\mu\rangle$ is an eigenstate of $\hat{p}$ defined by
(\ref{motrelB}), and where $V_\mu$ is an eigenvalue of the volume
operator corresponding to $V= |p|^{3/2}$ which  reads
\begin{equation}\label{motvol}
    \hat{V}|\mu\rangle = \Big( \frac{4 \pi \gamma |\mu|}{3}\Big)^{3/2}
    l_p^3\; |\mu\rangle =: V_\mu\,|\mu\rangle .
\end{equation}
The quantum operator corresponding to $H_g$ is defined to be
\cite{Ashtekar:2003hd,Ashtekar:2006uz}
\begin{equation}\label{mothamQ}
    \hat{H}_g := \hat{H}^{(\lambda)}_g \mid_{\lambda = \mu_o}
,~~~\mbox{~where}~~~0 < \mu_o \in \dR .
\end{equation}
Comparing (\ref{mothamQ}) with (\ref{mothamR}), and taking into account
(\ref{motcur}) one can see that the area of the square $\Box_{ij}$ is
not shrunk to zero, as required in the definition of the
classical curvature (\ref{motcur}), but determined at the
finite value of the area.

The mathematical justification proposed in
\cite{Ashtekar:2003hd,Ashtekar:2006uz} for such regularization  is
that one cannot define the local operator corresponding to
the curvature $F^k_{ab}$ because the 1-parameter group
$\hat{N}_\lambda$ is not weakly continuous at $\lambda =0$ in
$\mathcal{F}_g$ (dense subspace of $\mathcal{H}^g_{kin}$). Thus,
the limit $\lambda\,\rightarrow\, 0$ of $\hat{H}^{(\lambda)}_g$
does not exist.  To determine $\mu_o$ one proposes in
\cite{Ashtekar:2003hd,Ashtekar:2006uz,Ashtekar:2006wn} the
procedure which is equivalent to the following: We find  that the
area of the face of the cell $\mathcal{V}$ orthogonal to specific
direction is $Ar = |p|$. Thus the eigenvalue problem for the
corresponding kinematical operator of an area
$\widehat{Ar}:= |\hat{p}|$, due to (\ref{motrelB}), reads
\begin{equation}\label{motarea}
  \widehat{Ar}\,|\mu\rangle = \frac{4\pi \gamma l^2_p}{3}\,
  |\mu| \,|\mu\rangle =: ar (\mu)\,|\mu\rangle,~~~~\mu\in\dR ,
\end{equation}
where $ar (\mu)$ denotes the eigenvalue of $\widehat{Ar}$
corresponding to the eigenstate $|\mu\rangle$. On the other hand,
it is known that in LQG the kinematical area operator has
discrete eigenvalues \cite{Ashtekar:1996eg,Rovelli:1994ge}
and the smallest nonzero one, called an area gap $\Delta$, is
given by $\Delta = 2\sqrt{3}\,\pi \gamma l^2_p$. To identify
$\mu_o$ one postulates in \cite{Ashtekar:2006uz} that $\mu_o$ is
such that $ar (\mu_o) = \Delta$, which leads to $\mu_o =
3\sqrt{3}/2$. It is argued
\cite{Ashtekar:2003hd,Ashtekar:2006uz,Ashtekar:2006wn,Ashtekar:2007em}
that one cannot squeeze a surface to the zero value due to the
existence in the universe of the minimum quantum of area.
This completes the justification for the choice of the expression
defining the quantum Hamiltonian (\ref{mothamQ}) offered by LQC.

It is interesting to notice that for the model considered here
(defined on one-dimensional constant lattice) the existence of the
minimum area leads to the reduction of the non-separable space
$\mathcal{F}_g$ to its separable subspace. It is so because
due to (\ref{motrelC}) we have
\begin{equation}\label{motaction}
\hat{N}_{\mu_o}  \,|\mu\rangle =  |\mu + \mu_o\rangle,
\end{equation}
which means that the action of this operator does not lead outside
of the space spanned by $\{|\mu +k\,\mu_o\rangle, \,k\in \dZ\}$,
where $\mu \in \dR$ is fixed.

Finally, one can show (see, e.g.
\cite{Ashtekar:2003hd,Ashtekar:2006uz}) that the equation for
quantum dynamics, corresponding to (\ref{motker}), reads
\begin{equation}\label{motquanD}
B(\mu)\;\partial^2_\phi \psi(\mu,\phi) - C^+ (\mu)\psi (\mu +4
\mu_o , \phi) - C^- (\mu) \psi (\mu - 4 \mu_o, \phi) - C^0 (\mu)
\psi(\mu,\phi) = 0,
\end{equation}
where
\begin{equation}\label{motBC}
B(\mu):= \Big( \frac{2}{3 \mu_o}\Big)^6 \left[ |\mu + \mu_o|^{3/4}
- |\mu - \mu_o|^{3/4}\right]^6, \quad C^0 (\mu) := - C^+ (\mu) - C^-
(\mu),
\end{equation}
\begin{equation}\label{motCC}
C^+ (\mu):= \frac{\pi G}{9 |\mu_o|^3}\left|\;|\mu +3 \mu_o |^{3/2}-
|\mu +\mu_o |^{3/2}\right|, \quad C^-(\mu) := C^+ (\mu - 4 \mu_o ).
\end{equation}

Equation (\ref{motquanD}) has been derived formally by making use of
states which belong to $\mathcal{F}:= \mathcal{F}_g \otimes
\mathcal{F}_\phi$, where $\mathcal{F}_g$ and $\mathcal{F}_\phi$
are  dense subspaces of the kinematical Hilbert spaces
$\mathcal{H}^g_{kin}$ and $\mathcal{H}^\phi_{kin}$, respectively.
The space $\mathcal{F}$ provides an arena for the  derivation of
quantum dynamics. However, the physical states are expected
to be in $\mathcal{F}^\star $, the algebraic dual of $\mathcal{F}$
(see, e.g. \cite{Ashtekar:2003hd,Ashtekar:2006uz} and references
therein). It is known that $\mathcal{F}\subset \mathcal{H}_{kin}
\subset \mathcal{F}^\star $. Physical states are expected to have
the form $<\Psi|:= \sum_\mu \psi(\mu,\phi)<\mu|$, where $<\mu|$ is
the eigenbras of $\hat{p}$. One may give the structure of the
Hilbert space to some subspace of $\mathcal{F}^\star$ (constructed
from solutions to (\ref{motquanD})) by making use of the group
averaging method \cite{Marolf:2000iq,Ashtekar:1995zh} and obtain
this way the physical Hilbert space $\mathcal{H}_{phys}$.

The argument $\phi$ in $\psi(\mu,\phi)$ is interpreted as an
evolution parameter, $\mu$ is regarded as the physical degree of
freedom. Let us examine the role of the parameter $\mu_o$ in
(\ref{motquanD}). First of all, its presence causes that
(\ref{motquanD}) is a difference-differential equation so its
solution should be examined on a lattice. It is clear that some
special role must be played by $\mu_o = 0$ as the coefficient
functions of the equation, defined by (\ref{motBC}) and (\ref{motCC}),
are singular there. One can verify \cite{Ashtekar:2006uz} that as
$\mu_o \rightarrow 0$ the equation (\ref{motquanD}) turns into the
Wheeler-DeWitt equation
\begin{equation}\label{motwdw}
B(\mu)\;\frac{\partial^2}{\partial\phi^2} \psi(\mu,\phi)- \frac{16
\pi G}{3}\frac{\partial}{\partial\mu}\sqrt{\mu}\frac{\partial
}{\partial\mu}\,\psi(\mu,\phi) =
0,~~~~\mbox{with}~~~B(\mu):=\big|\frac{4\pi\gamma G
\hbar}{3}\,\mu\,\big|^{-3/2} .
\end{equation}

Equation (\ref{motquanD}) is not specially sensitive to any other
value of $\mu_o$. Thus, the determination of the numerical value
of this parameter by making use of the mathematical structure of
(\ref{motquanD}) seems to be impossible.

\subsection{Minimum length problem}

The singularity resolution offered by LQC, in the context of flat
FRW universe, is a striking result.  Let us look at the key
ingredients  of the construction of LQC which are responsible for
this long awaited result:

Discussing the mathematical structure of the constraint equation
we have found that $\mu_o$ must be a non-zero  if we wish to deal
with the regular (\ref{motquanD}) instead  of the singular
(\ref{motwdw}). However, the numerical value of $\mu_o$ cannot be
determined from  the equation (\ref{motquanD}). It plays the role of
a free parameter if it is not specified.

The parameter $\mu_o$ enters the formalism due to the
representation of the curvature of the connection $F^k_{ab}$ via
the holonomy around a loop (\ref{motcur}). The smaller the loop
the better approximation we have. The size of the loop, $\mu_o$,
determines the quantum operator corresponding to the modified
gravitational part of the Hamiltonian (\ref{mothamQ}). One may
determine $\mu_o$ by making use of an area of the loop (used
in fact as a technical tool). Thus, the spectrum of the
quantum operator corresponding to an area operator,
$\widehat{Ar}$, seems to be a suitable source of information on
the possible values of $\mu_o$. Previous section shows explicitly that
the construction of the quantum level is heavily  based on the
kinematical ingredients of the formalism. Thus, it is
natural to explore the kinematical $\widehat{Ar}$ of LQC. However,
its spectrum (\ref{motarea}) is continuous so it is useless for
the determination of $\mu_o$. On the other hand, the spectrum of
kinematical $\widehat{Ar}$ of LQG is discrete
\cite{Ashtekar:1996eg,Rovelli:1994ge}. Thus, it was tempting to
use such a spectrum to fix $\mu_o$ postulating that the  minimum
quantum of area defines the minimum area of the loop
defining  (\ref{mothamQ}). This way $\mu_o$ has been fixed.

The physical justification, however, for such procedure is doubtful because LQC is not the cosmological sector of LQG. Therefore, Eq. (\ref{mothamQ}) includes an insertion by hand of specific properties of the spectrum of $\widehat{Ar}$ from LQG into LQC \cite{Bojowald:2008ik}. After all, the area gap of the spectrum of $\widehat{Ar}$ of LQG is not a fundamental constant (like the speed of light, Planck's constant, Newton's constant) so its use in the context of LQC has poor physical justification.

\subsection{Summary}

First of all we have shown that the introduction of the quantum of an area from LQG into LQC at kinematical level is only an assumption. As a consequence, the energy scale of the Big Bounce described by $\lambda$ parameter is in fact unknown. It is so because $\lambda$ is a free parameter of LQC.

\bigskip

\goodbreak
\section{Formalism of nonstandard LQC}

Now we will present an alternative method of canonical quantization of cosmological models of GR, which makes use of loop geometry \cite{Dzierzak:2009ip,Malkiewicz:2009qv, Mielczarek:2010rq}. We believe that the nonstandard LQC may be related with the reduced phase space quantization of Loop Quantum Gravity \cite{Giesel:2007wn}. What is the motivation for developing an alternative formalism? First of all, agreement of results obtained with both methods would be a sort of proof that the procedure of quantization is correct. Of course the final test is always an agreement with observational data when they become available. Another reason to develop an alternative approach is improving our understanding of some conceptual issues like identification of physical observables or quantum evolution of a system with the Hamiltonian constraint.

\subsection{Main idea}

In the nonstandard LQC \cite{Dzierzak:2009ip,Malkiewicz:2009qv} one first solves the constraint (the constraints) at the classical level to identify the physical phase space (i.e. the space of Dirac's observables).
Secondly, in that space one finds the elementary observables and their algebra. These elementary observables are used as ``building blocks'' for the compound observables, like the energy density or the volume of the universe, so they have deep physical meaning.
The compound observables are thus defined on the physical phase space too.
Thus, their properties may be confronted in future with the data of observational cosmology.
The compound observables depend on the elementary ones and an evolution parameter\footnote{which is value of the scalar field $\phi$}, so for fixed moments of time they are functions only of elementary observables. Next step is a quantization. By this term we mean finding a self-adjoint representation of the algebra of the elementary observables and solution to the eigenvalue problem for operators corresponding to the compound observables \cite{Malkiewicz:2009qv}.

The difference of understanding the term ``quantization'' is a source of another difference between these two LQC methods. In nonstandard formalism, approximation of the curvature of connection by a holonomy along a loop of finite size (modification of Hamiltonian by loops) is done entirely at the classical level. Our approach is different from the so-called polymerization method practisized by users of standard LQC and treated as an effective quantum theory (see, e.g. \cite{Chiou:2007mg}), where the modification in the Hamiltonian: $\beta \rightarrow \sin(\lambda \beta)/\lambda$ finishes the procedure of quantization. It means that in our method resolution of the singularities happens at the classical level due to loop modification of Hamiltonian. This modification is parameterized by a continuous parameter $\lambda $. There is no specific choice of $\lambda$, so we can say that $\lambda$ is a free parameter. Details on theoretical ways of finding the value of $\lambda$ can be found in conclusions of this Thesis.

There is also another important issue. Why should we quantize a cosmological model which is free from the cosmological singularity? We have at least three reasons:
(i) to make comparison with the standard LQC results, we must have a quantum model; (ii) the parameter $\lambda$ specifying the modification is a free parameter in nonstandard LQC. As the result, the critical density of matter at the bounce becomes unspecified as it depends on $\lambda$. Since it may become arbitrarily big for small enough $\lambda$, the system may enter an arbitrarily small length scale, where quantum effects cannot be ignored \cite{Dzierzak:2009ip}; (iii) making predictions of our model for quantum cosmic data may be used to fix the free parameter $\lambda$, after such data become available.

\subsection{Comparison of both LQC methods}

Shortly, one can write:

\noindent the standard LQC = first quantize, then impose constraints = Dirac's method;

\noindent the nonstandard LQC = first solve constraints, then quantize = reduced phase space quantization.

The most important advantage of the nonstandard LQC is that the spectra of the operators are directly obtained on physical Hilbert space. In the case of the standard LQC, one firstly obtains results on the kinematical Hilbert space. The physical states are obtained using kernels of quantum constraints operators. Applying group averaging metods leads to physical spectra of observables. Thus our nonstandard method is simpler and more efficient than the standard one.

Another important feature of nonstandard LQC is that this method is fully controlled analytically as it does not require any numerical work, at least in FRW and Bianchi I cases, in contrast to the standard LQC results.

In the nonstandard LQC an evolution parameter $\phi$ stays classical during the quantization. This happens because $\phi$ does not belong to the physical phase space. In the standard LQC $\phi$ is a phase space variable and should be quantized. This is crucial because $\phi$ being a quantum variable may fluctuate, which may makes an interpretation of $\phi$ problematic.

\bigskip

\goodbreak
\section{Application of nonstandard LQC to FRW}

In this chapter we consider, as an application of our method, the simplest cosmological model, namely FRW model with $k=0$ and with free massless scalar field $\phi$ in space with topology $\dR^3$.

\subsection{Hamiltonian}

In what follows we use the ``improved'' scheme $\bar\mu= \sqrt{\frac{1}{|p|}}\,\lambda$ \cite{Ashtekar:2006wn}. Using it and taking (\ref{mothol}) we calculate (\ref{mothamL}) and get the modified total Hamiltonian $H^{(\lambda)}_g$ corresponding to (\ref{motham}) in the form
\begin{equation}\label{frwregH}
  H^{(\lambda)}/N = -\frac{3}{8\pi G \gamma^2}\;\frac{\sin^2(\lambda
\beta)}{\lambda^2}\;v + \frac{p_{\phi}^2}{2\, v},
\end{equation}
where
\begin{equation}\label{frwre1}
    \beta := \frac{c}{|p|^{1/2}},~~~v := |p|^{3/2}
\end{equation}
are the canonical variables proposed in \cite{Ashtekar:2006wn}.
The variable $\beta = \gamma\dot{a}/a$ so it corresponds to the
Hubble parameter $\dot{a}/a$, whereas $v^{1/3} = a$ is
proportional to the scale factor $a$.

The complete Poisson bracket for the canonical variables
$(\beta,v,\phi,p_\phi)$ is defined to be
\begin{equation}\label{frwre2}
    \{\cdot,\cdot\}:= 4\pi G\gamma\;\bigg[ \frac{\partial \cdot}
    {\partial \beta} \frac{\partial \cdot}{\partial v} -
     \frac{\partial \cdot}{\partial v} \frac{\partial \cdot}{\partial \beta}\bigg] +
     \frac{\partial \cdot}{\partial \phi} \frac{\partial \cdot}{\partial p_\phi} -
     \frac{\partial \cdot}{\partial p_\phi} \frac{\partial \cdot}{\partial
     \phi}.
\end{equation}
The dynamics of a canonical variable $\xi$ is defined by
\begin{equation}\label{frwdyn}
    \dot{\xi} := \{\xi,H^{(\lambda)}\},~~~~~~\xi \in \{\beta,v,\phi,p_\phi\},
\end{equation}
where $\dot{\xi} := d\xi/d\tau$, and where $\tau$ is an evolution
parameter. The dynamics in the physical phase space,
$\mathcal{F}_{phys}^{(\lambda)}$, is defined by solutions to
(\ref{frwdyn}) satisfying the condition $H^{(\lambda)}\approx 0$. The
solutions of (\ref{frwdyn}) ignoring the constraint
$H^{(\lambda)}\approx 0$ are in the kinematical phase
space, $\mathcal{F}_{kin}^{(\lambda)}$.

\subsection{Classical dynamics}

Equation (\ref{frwregH}) can be rewritten as
\begin{equation}\label{product}
  H^{(\lambda)} = N\,H_0^{(\lambda)}\,\tilde{H}^{(\lambda)}\approx 0,
\end{equation}
where
\begin{equation}\label{frwdefprod}
H_0^{(\lambda)} := \frac{3}{8 \pi G \gamma^2 v} \;\Big(\kappa
\gamma |p_\phi| + v\,\frac{|\sin(\lambda
\beta)|}{\lambda}\Big),~~~~~~ \tilde{H}^{(\lambda)}:= \kappa
\gamma |p_\phi| - v\, \frac{|\sin(\lambda \beta)|}{\lambda},
\end{equation}
where $\kappa^2 \equiv 4\pi G/3$.

\noindent It is clear that $H_0^{(\lambda)} = 0$ only in the case
when $p_\phi =0=\sin(\lambda \beta)$. Such case, due to
(\ref{frw1a})-(\ref{frw5a}), implies no dynamics.\\
Choosing the gauge $N := 1/H_0^{(\lambda)}$ (which simplifies the calculations) we get
\begin{eqnarray}\label{frw1a}
  \dot{p_{\phi}}&=&0, \\\ \label{frw2a}
  \dot{\beta}&=&  -4\pi G\gamma
  \;\frac{|\sin(\lambda\, \beta)|}{\lambda},
  \\ \label{frw3a}
  \dot{\phi}&=& \kappa\gamma~\textrm{sgn}(p_{\phi}), \\ \label{frw4a}
  \dot{v}&=& 4\pi G\gamma v \cos(\lambda\,
  \beta)~\textrm{sgn}(\sin(\lambda\, \beta)), \\ \label{frw5a}
  \tilde{H}^{(\lambda)} & = & 0.
\end{eqnarray}

Combining (\ref{frw3a}) with (\ref{frw4a})  gives
\begin{eqnarray}\label{frwpp1}
\frac{\dot v}{\dot \phi} =  3\kappa v \cos{(\lambda
\beta)}\;\textrm{sgn}(\sin{(\lambda
\beta)})\;\textrm{sgn}(p_{\phi}).
\end{eqnarray}
Rewriting (\ref{frwpp1}) (and using $\dot v/\dot \phi = dv/d\phi$)
gives
\begin{eqnarray}\label{frwzcos}
\frac{\textrm{sgn}(\sin(\lambda \beta))}{\cos(\lambda
\beta)}\;\frac{dv}{v} = 3\kappa\;\textrm{sgn}(p_{\phi})\;d\phi
\end{eqnarray}
Making use of the identity $\;\sin^2(\lambda \beta)+\cos^2(\lambda
\beta)=1$ and (\ref{frw5a}) gives
\begin{eqnarray}\label{frwnacos}
|\cos{(\lambda \beta)}|= \sqrt{1-\Big(\frac{\kappa \gamma p_{\phi}
\lambda }{v}}\Big)^2
\end{eqnarray}
Combining (\ref{frwzcos}) with (\ref{frwnacos}), for $\beta\in ]0, \pi/2
\lambda[$, leads to
\begin{equation}\label{frwdynamika}
    \frac{dv}{\sqrt{
   v^2-(\kappa\gamma\lambda
   p_\phi)^2}}= 3 \kappa\; \textrm{sgn}(p_{\phi})\;d\phi .
\end{equation}
Since $p_\phi$ is just a constant (due to (\ref{frw1a})) we can
easily integrate (\ref{frwdynamika}) and get
\begin{equation}\label{frwint}
    \ln\bigg|v+\sqrt{v^2-(\kappa\gamma\lambda
   p_\phi)^2}\bigg|=3 \kappa\;\textrm{sgn}(p_{\phi})(\phi-\phi_0) .
\end{equation}
Rewriting (\ref{frwint}) leads to
\begin{equation}\label{frwres2}
2\,v = \exp{\big(3\kappa\;\textrm{sgn}(p_{\phi})\;(\phi -
\phi_{0})\big)} + (\kappa\gamma |p_{\phi}|
\lambda)^2\cdot\exp{\big(-3\kappa\;\textrm{sgn}(p_{\phi})\;(\phi -
\phi_{0})\big)}.
\end{equation}
The solution for the variable $\beta$ may be easily determined
from (\ref{frw5a}) rewritten as
\begin{equation}\label{frwres3}
\kappa \gamma |p_\phi| = v\, \frac{|\sin(\lambda \beta)|}{\lambda}
\end{equation}
Finally we get
\begin{equation}\label{frwres4}
    \sin(\lambda \beta)= \frac{2 \kappa\gamma\lambda
   |p_\phi|}{\exp\big(3\kappa\, \textrm{sgn}(p_{\phi})\,(\phi-\phi_0)\big)+
   (\kappa\gamma\lambda
   p_\phi)^2\exp \big(-3\kappa\, \textrm{sgn}(p_{\phi})\,(\phi-\phi_0)\big)}
\end{equation}
where the domain of the variable $\beta$ has been extended to the
interval $ ]0,\pi/\lambda[$.

Equations (\ref{frwres2}) and (\ref{frwres4}) present the dependence of
the canonical variables $v$ and $\beta$ on the evolution parameter
$\phi$, which is a monotonic function due to (\ref{frw3a}).

\subsection{Observables}

\subsubsection{Elementary observables and their algebra}

A function, $\mathcal{O}$, defined on phase space is a Dirac
observable if
\begin{equation}\label{frwdirac}
\{\mathcal{O},H^{(\lambda)}\} \approx 0.
\end{equation}
Since we have
\begin{equation}\label{frwDirac}
\{\mathcal{O},H^{(\lambda)}\}=  \{\mathcal{O},N H_0^{(\lambda)}
\tilde{H}^{(\lambda)}\}= N H_0^{(\lambda)}\{\mathcal{O},
\tilde{H}^{(\lambda)}\} + \{\mathcal{O}, N
H_0^{(\lambda)}\}\tilde{H}^{(\lambda)},
\end{equation}
it is clear that on the constraint surface,
$\tilde{H}^{(\lambda)}=0$, the Dirac observable satisfies
(independently on the choice of $N$) a much simpler equation
\begin{equation}\label{frwdonot}
\{\mathcal{O},\tilde{H}^{(\lambda)}\}\approx 0.
\end{equation}
Using the gauge $N:= 1/H_0^{(\lambda)}$ and solve (\ref{frwdirac}) in
the whole phase space, i.e. we solve the equation
\begin{equation}\label{frwdir}
\frac{\sin(\lambda\beta)}{\lambda}\,\frac{\partial
\mathcal{O}}{\partial\beta} - v \cos(\lambda\beta)\,\frac{\partial
\mathcal{O}}{\partial v} - \frac{\kappa\,\textrm{sgn}(p_{\phi})}{4
\pi G}\,\frac{\partial \mathcal{O}}{\partial\phi} = 0.
\end{equation}

A function $\mathcal{O} =
\mathcal{O}(\mathcal{O}_1,\ldots\mathcal{O}_k)$ satisfies
(\ref{frwdir}) if
\begin{equation}\label{frwobser}
\{\mathcal{O}_1,\tilde{H}^{(\lambda)}\} = 0=
\{\mathcal{O}_2,\tilde{H}^{(\lambda)}\} =\ldots =
\{\mathcal{O}_k,\tilde{H}^{(\lambda)}\},
\end{equation}
where $k+1$ is the dimension  of the kinematical phase
space. It is so because one has
\begin{equation}\label{frwprop}
\{\mathcal{O},\tilde{H}^{(\lambda)}\} = \frac{\partial
\mathcal{O}}{\partial
\mathcal{O}_1}\,\{\mathcal{O}_1,\tilde{H}^{(\lambda)}\}  + \ldots
+ \frac{\partial \mathcal{O}}{\partial
\mathcal{O}_k}\,\{\mathcal{O}_k,\tilde{H}^{(\lambda)}\}.
\end{equation}

In what follows we consider only elementary observables. The
set of such observables, $\mathcal{E}$, is defined by the
requirements: (i) each element of $\mathcal{E}$ is  a solution to
(\ref{frwdir}), (ii) elements of $\mathcal{E}$ are functionally
independent on the constraint surface, $\tilde{H}^{(\lambda)}=0$,
(iii) elements of $\mathcal{E}$ satisfy a Lie algebra, and (iv)
two sets of observables satisfying two algebras are considered to
be the same if these algebras are isomorphic.

In our case $k = 3$ and  solutions to (\ref{frwdir}) are found to be
\begin{equation}\label{frwobser1}
\mathcal{O}_1:= p_{\phi},~~~\mathcal{O}_2:= \phi -
\frac{s}{3\kappa}\;\textrm{arth}\big(\cos(\lambda \beta)\big),~~~~
\mathcal{O}_3:= s\,v\, \frac{\sin(\lambda \beta)}{\lambda},
\end{equation}
where $s := \textrm{sgn}(p_\phi)$. One may verify that the
observables satisfy the Lie algebra
\begin{equation}\label{frwala1}
\{\mathcal{O}_2,\mathcal{O}_1\}=
1,~~~~\{\mathcal{O}_1,\mathcal{O}_3\}= 0,~~~~
\{\mathcal{O}_2,\mathcal{O}_3\}=  \gamma\kappa .
\end{equation}

Because of the constraint $\tilde{H}^{(\lambda)}=0$ (see
(\ref{frwres3})), we have
\begin{equation}\label{frwcon}
\mathcal{O}_3=  \gamma \kappa \,\mathcal{O}_1.
\end{equation}
Thus,  we have only two elementary Dirac observables which may be
used to parameterize  the physical phase space
$\mathcal{F}_{phys}^{(\lambda)}$. To identify the Poisson bracket
in $\mathcal{F}_{phys}^{(\lambda)}$ consistent with the Poisson
bracket (\ref{frwre2}) defined in $\mathcal{F}_{kin}^{(\lambda)}$, we
find a symplectic twoform corresponding to (\ref{frwre2}). It reads
\begin{equation}\label{frwsym1}
    \omega = \frac{1}{4 \pi G \gamma} d\beta \wedge d v + d\phi
    \wedge dp_\phi .
\end{equation}
The twoform $\omega$ is degenerate on
$\mathcal{F}_{phys}^{(\lambda)}$ due to the constraint
$\tilde{H}^{(\lambda)}=0$. Making use of the explicit form of this
constraint (\ref{frwres3}) and  the functional form of
$\mathcal{O}_1$ and $\mathcal{O}_2$, leads to the symplectic form
$\Omega$ on $\mathcal{F}_{phys}^{(\lambda)}$. Direct calculations
give
\begin{equation}\label{frwsym2}
    \Omega := \omega_{| \tilde{H}^{(\lambda)}=0} = d\, \mathcal{O}_2
    \wedge d\, \mathcal{O}_1 ,
\end{equation}
where $\omega_{| \tilde{H}^{(\lambda)}=0}$ denotes the reduction
of $\omega$ to the constraint surface. The Poisson bracket
corresponding to (\ref{frwsym2}) reads
\begin{equation}\label{frwsym3}
\{\cdot,\cdot\}:=\frac{\partial\cdot}{\partial
\mathcal{O}_2}\frac{\partial\cdot}{\partial \mathcal{O}_1} -
\frac{\partial\cdot}{\partial
\mathcal{O}_1}\frac{\partial\cdot}{\partial \mathcal{O}_2}
\end{equation}
so the algebra satisfied by $\mathcal{O}_1$ and $\mathcal{O}_2$
has a simple form given by
\begin{equation}\label{frwsym4}
\{\mathcal{O}_2,\mathcal{O}_1\}= 1.
\end{equation}

Our kinematical phase space, $\mathcal{F}_{kin}^{(\lambda)}$, is
four dimensional. In relative dynamics one variable is used
to parameterize three others. Since the constraint relates  the
variables, we have only two independent variables. This is the
reason  we have only two elementary physical observables
parameterizing $\mathcal{F}_{phys}^{(\lambda)}$.

\subsubsection{Functions on phase space}

Now we discuss the functions on the constraint
surface that may describe singularity aspects of our cosmological
model. Considered functions are functions of elementary observables and evolution parameter $\phi$, so they are not observables. They become observables for fixed $\phi$.

An interesting function is energy density $\rho$ of the scalar field
$\phi$
\begin{equation}\label{frwrho2}
\rho(\lambda,\phi)=\frac{1}{2}\,\frac{p_{\phi}^2}{v^2}.
\end{equation}
In terms of elementary observables we have
\begin{equation}\label{frwobser2}
p_\phi = \mathcal{O}_1,~~~~v = \kappa\gamma\lambda\,
    |\mathcal{O}_1|\,\cosh\big(3\kappa  (\phi- \mathcal{O}_2)\big)
\end{equation}
which means that
\begin{equation}\label{frwrho3}
\rho(\lambda,\phi)= \frac{1}{2}\,\frac{1}{(\kappa\gamma\lambda)^2\,
    \cosh^2 3\kappa  (\phi- \mathcal{O}_2)}.
\end{equation}

For fixed $p_\phi$ the density $\rho$ takes its maximum value at
the minimum value of $v$. Rewriting (\ref{frwres2}) in the form
\begin{equation}\label{id1}
\frac{v}{\triangle}= \cosh\big(3 \kappa s (\phi - \phi_{0}) - \ln
\triangle\big),~~~~~\mbox{where}~~~~~\triangle:=
\kappa\gamma\lambda\,|p_\phi|,
\end{equation}
we can see that  $\,\cosh(\cdot)\,$  takes minimum value equal to
one at $ 3 \kappa s\,(\phi - \phi_{0})= \ln \triangle$. Thus, the
maximum value of the density, $\rho_{\max}$, corresponds to $v =
\triangle$ and  reads
\begin{equation}\label{frwcr1}
\rho_{\max} = \frac{1}{2\kappa^2 \gamma^2}\,\frac{1}{ \lambda^2}.
\end{equation}
which means that the Big Bounce occurs at the classical level due to modification classical Hamiltonian by loops.
We can determine $\rho_{\max}$ if we know $\lambda$, but as we now $\lambda$ is a free parameter of the formalism.

Another interesting observable is a volume. In terms of observables it reads
\cite{Dzierzak:2009ip}
\begin{equation}\label{frwvol}
    v(\lambda,\phi) = \kappa\gamma\lambda\,
    |\mathcal{O}_1|\,\cosh3\kappa  (\phi-
    \mathcal{O}_2).
\end{equation}

Looking at the equations written above one can see that $\rho$ and $V$ depend explicitly on observables $\mathcal{O}_1$ and $\mathcal{O}_2$ and evolution parameter $\phi$ so they are observables for fixed value of $\phi$.

\subsection{Quantization}

As we remember, apart from resolving the singularity there are some important reasons to continue the procedure of the quantization.

\subsubsection{Representation of elementary observables}

For the classical algebra (\ref{frwsym4}) two representations are used
\begin{equation}\label{frwquant1}
\mathcal{O}_1 \longrightarrow \widehat{\mathcal{O}}_1 f(x):=
-i\,\hbar\,\partial_x f(x),~~~~\mathcal{O}_2 \longrightarrow
\widehat{\mathcal{O}}_2 f(x):= \widehat{x} f(x) := x f(x),
\end{equation}
which leads to $[\widehat{\mathcal{O}}_1,\widehat{\mathcal{O}}_2]=
- i\,\hbar\,\id$, and
\begin{equation}\label{frwquant2}
\mathcal{O}_1 \longrightarrow \widehat{\mathcal{O}}_1 f(x):=
\widehat{x} f(x) := x f(x),~~~~\mathcal{O}_2 \longrightarrow
\widehat{\mathcal{O}}_2 f(x):= -i\,\hbar\,\partial_x f(x),
\end{equation}
which leads to $[\widehat{\mathcal{O}}_1,\widehat{\mathcal{O}}_2]=
i\,\hbar\,\id$, where $x\in \mathbb{R}$.\\ Due to the Stone$-$von
Neumann theorem all self-adjoint representations of the algebra
(\ref{frwsym4}) are unitarily equivalent to the representation
(\ref{frwquant1}) or (\ref{frwquant2}) defined on a suitable dense
subspace of $L^2(\dR)$. In that sense the choice of representation is unique.

\subsubsection{Energy density operator}

The representation (\ref{frwquant2}) is essentially self-adjoint on
the dense subspace $D$ of the Hilbert space $L^2 [-r,r],\,$
where $r \in \dR_+$, defined to be
\begin{equation}\label{frwden}
D:= \{f\in C^\infty [-r,r]\,|\, f^{(n)}(-r) = f^{(n)}(r),\, n \in
\{0\}\cup \dN \},
\end{equation}
where $f^{(n)} := d^{\,n}f/dx^n$.\\ The eigenvalue problem,
$\widehat{\mathcal{O}}_2 f_p = p\, f_p$, has the solution
\begin{equation}\label{frwsol1}
f_p (x) = (2r)^{-1/2} \exp(i x p /\hbar),~~~~p(k):= 2\pi\hbar
k/r,~~k \in \dZ .
\end{equation}
The spacing of neighboring eigenvalues $\square$ is defined by
\begin{equation}\label{frwspa}
\square := p(k+1) - p(k) = 2\pi\hbar/r
\end{equation}
Making $r$ sufficiently large $\square$ can be made as small as desired, which means that the spectrum of $\widehat{\mathcal{O}}_2 $ is continuous.

In the representation (\ref{frwquant2}) the energy density operator reads
\begin{equation}\label{frwquant3}
 \widehat{\rho}:=\frac{1}{2}
    \frac{1}{(\kappa\gamma\lambda)^2\cosh^2 3\kappa(\phi+i\,\hbar\,\partial_x)}.
\end{equation}
Since $\widehat{\mathcal{O}}_2$ is essentially self-adjoint
on $\mathcal{F}_r := \{f_{p(k)}\}_{k \in \dZ}$, we may apply the
spectral theorem to get $\widehat{\rho}\,f_p = \rho (\phi,\lambda,p)\,f_p$ where
\begin{equation}\label{frwquant5}
\rho (\phi,\lambda,p) := \frac{1}{2} \frac{1}{(\kappa\gamma
\lambda)^2\cosh^2 3\kappa(\phi - p)},
\end{equation}
and where $\rho (\phi,\lambda,p)$ is the eigenvalue corresponding
to the eigenvector $f_p$.

It is clear from our results that classical (\ref{frwrho2}) and quantum
(\ref{frwquant5}) expressions for the energy density coincide.
One may verify that the maximum density $\rho_{max}(\lambda) = \frac{1}{2}\frac{1}{(\kappa\gamma
\lambda)^2}$.

Starting from the other representation we would get the quantum model of the energy
density presented in \cite{Malkiewicz:2009zd}, which is equivalent
this present one.

\subsubsection{Volume operator}

To define the quantum operator corresponding to $v$, we introduce $w$ defined by
\begin{equation}\label{frwvol1}
w := \kappa\gamma\lambda\;\mathcal{O}_1\;\cosh3\kappa(\phi-
\mathcal{O}_2).
\end{equation}
Since $v = |w|$ it is clear that quantization of $v$ reduces to the quantization of $w$. The latter may be done in a standard way as
follows
\begin{equation}\label{frwvol2}
\hat{w}\,f(x) :=
    \kappa\gamma\lambda\,\frac{1}{2}\,\bigg(
    \widehat{\mathcal{O}}_1\,\cosh3\kappa  (\phi-
    \widehat{\mathcal{O}}_2)+
     \cosh3\kappa  (\phi-
    \widehat{\mathcal{O}}_2)\;\widehat{\mathcal{O}}_1\bigg) f(x),
\end{equation}
where $f \in  L^2 (\dR)$.\\ For the elementary observables
$\mathcal{O}_1$ and $\mathcal{O}_2$ we use the  representation
(\ref{frwquant1}). An explicit form of the operator $\hat{w}$ reads
\begin{equation}\label{frwrepp1}
\hat{w}\,f(x)= i\,\frac{\kappa\gamma\lambda\hbar}{2}\bigg(
    2 \cosh3\kappa(\phi-x)\;\frac{d}{dx}
     -3\kappa\sinh3\kappa
    (\phi-x)\bigg)\,f(x).
\end{equation}
To simplify our considerations we take $f$ in the form
\begin{equation}\label{frweq3}
f(x) :=  A\,e^{i h(x)}\, \cosh^{-1/2} 3 \kappa (\phi - x),
\end{equation}
where $h$ is a real-valued function and $A \in \dR $.

\subsubsection{Eigenvalue problem}

Considering the eigenvalue problem for the operator $\hat{w}$
in the set of functions of the form (\ref{frweq3}). We get
\begin{equation}\label{frweq4}
\hat{w}\, f (x)= -\kappa\gamma\lambda\hbar
\cosh3\kappa(\phi-x)\frac{d h(x)}{dx}\, f(x) =: b\, f (x),
\end{equation}
where  $b \in \dR $ is the eigenvalue of $\hat{w}$.\\ A general form of $h$ satisfying (\ref{frweq4}) is given by
\begin{equation}\label{frweq55}
 h (x)=  \frac{2  b}{3\kappa^2
    \gamma\lambda\hbar}\arctan
    e^{3\kappa(\phi-x)},
\end{equation}
which means that a normalized $f_b $ satisfying (\ref{frweq4}) reads
\begin{equation}\label{frweq5}
f_b (x):= \frac{\sqrt{\frac{3\kappa}{\pi}}\exp\Big(i \frac{2
b}{3\kappa^2 \gamma\lambda\hbar}\arctan
    e^{3\kappa(\phi-x)}\Big)}{\cosh^{\frac{1}{2}}3\kappa(\phi-x)}.
\end{equation}

\subsubsection{Orthogonality}

Using (\ref{frweq5}) we get
\begin{equation}
    \langle f_b|f_a\rangle= \frac{3\kappa}{\pi}\int_{-\infty}^{\infty}\frac{\exp\Big(i \frac{2
(a-b)}{3\kappa^2 \gamma\lambda\hbar}\arctan
    e^{3\kappa(\phi-x)}\Big)}{\cosh3\kappa(\phi-x)}~dx .
\end{equation}

The substitution $\tan z =e^{3\kappa(\phi-x)}$ leads to
\begin{equation}
\langle f_b|f_a\rangle=\frac{2}{\pi}\int_{0}^{\frac{\pi}{2}}\exp\Big(i \frac{2
(a-b)}{3\kappa^2 \gamma\lambda\hbar}z\Big)~dz=-i\,
\frac{3\kappa^2\gamma\lambda\hbar}{\pi(a-b)}\exp\Big(i \,\frac{2
(a-b)}{3\kappa^2 \gamma\lambda\hbar}z\Big)\Big|_0^{\frac{\pi}{2}}.
\end{equation}
One may verify that $\langle f_b|f_a\rangle = 0\;$ iff
\begin{equation}\label{frwab}
    a-b = 6\kappa^2\gamma\lambda\hbar\,m = 8\pi
    G\gamma\lambda\hbar\,m,~~~~m\in \mathbb{Z} .
\end{equation}
Thus, the set $\mathcal{F}_b:=\{~f_a\;|\; a = b + 8\pi G\gamma\lambda\hbar\,m;~m\in \dZ;~b \in \dR~\}$ is orthonormal. Each subspace $\mathcal{F}_b \subset L^2(\dR)$ spans a pre-Hilbert space. The completion of each $span \;\mathcal{F}_b,~~\forall b \in \dR$, gives $L^2(\dR)$ in the norm of $L^2(\dR)$.

\subsubsection{Self-adjointness}

The operator $\hat{w}$ is symmetric on $\mathcal{F}_b $ for any $b \in\dR$ due to $\langle f_b|\hat{w}f_a\rangle - \langle\hat{w}f_b|f_a\rangle = (a-b)\langle f_b|f_a\rangle$ because $\langle f_b|f_a\rangle=0$ for $a\neq b$.

To examine the self-adjointness of the unbounded operator
$\hat{w}$, we first identify the deficiency subspaces of this operator
$\mathcal{K_\pm}$ \cite{RS,DS}
\begin{equation}\label{frwself1}
\mathcal{K_\pm}:= \{g_\pm \in D_b (\hat{w}^\ast)~|~\langle
g_\pm|(\hat{w} \pm i \id)f_a \rangle =0, \;\forall f_a \in D_b
(\hat{w})\},
\end{equation}
where $D_b (\hat{w}):=$ {\it span} $\mathcal{F}_b$, and $D_b(\hat{w}^{\ast}):=\{f\in L^2(\dR) :~\exists !f^{\ast}~\langle
f^{\ast}|g\rangle=\langle f|\hat{w}g\rangle,~\forall g\in
D_b(\hat{w})\}$.

For each $f_a \in D_b(\hat{w})\subset L^2(\dR)$ we have
\begin{equation}\label{frwself2}
0 = \langle g_\pm|(\hat{w} \pm i \id)f_a\rangle = (a \pm i)
\int_{-\infty}^\infty dx \;\overline{g_\pm (x)} f_a
(x)~~~~~\Longrightarrow~~~~~g_+ = 0 = g_- .
\end{equation}
Thus, the deficiency indices $n_\pm := dim [\mathcal{K}_\pm]$ of
$\hat{w}$ satisfy the relation: $n_+ = 0 = n_-$. This proves essential self-adjointness of $\hat{w}$ on $D_b(\hat{w})$.

\subsubsection{Spectrum}

Due to the spectral theorem on self-adjoint operators
\cite{RS,DS}, we may quantize the volume as
follows
\begin{equation}\label{frwsp1}
 v = |w|~~~\longrightarrow~~~\hat{v} f_a :=  |a| f_a .
\end{equation}

A common feature of all $\mathcal{F}_b$ is the existence of the minimum gap $\bigtriangleup:= 8\pi G\gamma\hbar\,\lambda\;$ defining a quantum of the volume. Let us discuss this issue in more detail. Denoting the minimum eigenvalue of $\hat{v}$ by $v_{min}$, one can verify that $v_{min}=\textrm{min}\{b,\Delta-b\}$, where $b\in [0,\Delta[$. The spectrum consists of the union of $\{v_{min}+n\Delta\}$ and
$\{-v_{min}+(n+1)\Delta\}$, where $n=0,1,\dots$ There are only two cases when these two subsets are identical, namely when $\,v_{min}=0\,$ or $\,v_{min}=\Delta/2\,$, for which the minimum gap $\Delta$ is a constant gap between any two adjacent levels of the spectrum. Otherwise, the gap equals either $\,\Delta - 2v_{min}\,$ or $\,2v_{min}\,$, and the minimum gap is the smaller one. One can verify that the case of any $b \in \dR$ reduces to the above case.

There is no quantum of the volume in the limit $\lambda \rightarrow 0$, corresponding to the classical FRW model without the loop geometry modification.

It results from (\ref{frwab}) that for $b=0$ and $m=0$ the minimum eigenvalue of $\hat{v}$ equals zero. This special case corresponds to the classical situation when $v=0$, which due to (\ref{frwregH}) means that $p_\phi = 0$ (no classical dynamics). Thus, we
have a direct correspondence between classical and quantum levels corresponding to this very special state. All other states describe bouncing dynamics.

\subsubsection{Evolution}

It is clear that the relation between eigenvectors corresponding to the same
eigenvalue for different values of the parameter $\phi$ reads $f_a^{\phi+\psi} = e^{\psi\partial_\phi}f_a^{\phi} =e^{-i\frac{\psi}{\hbar} \widehat{\mathcal{O}}_1}f_a^{\phi}$.

One may verify that
\begin{equation}\label{frwda1}
    \hat{w}(\phi+\psi)=\cosh{(3\kappa\psi)}\hat{w}(\phi)
    +\frac{\sinh{(3\kappa\psi)}}
    {3\kappa}\partial_{\phi}\hat{w}(\phi),
\end{equation}
thus
\begin{eqnarray}\nonumber
    &&\langle f_b^{\phi}|\hat{w}(\phi+\psi)f_a^{\phi}\rangle =
    \langle f_b^{\phi}|\hat{w}(\phi)f_a^{\phi}\rangle\cosh{(3\kappa\psi)}+
    \frac{\sinh{(3\kappa\psi)}}{3\kappa}\langle f_b^{\phi}|\partial_{\phi}
    \hat{w}(\phi)f_a^{\phi}\rangle\\&&=a\cosh{(3\kappa\psi)}\,
    \delta_{ab}+(b-a)\frac{\sinh{(3\kappa\psi)}}{3\kappa}\langle
    f_b^{\phi}|\partial_xf_a^{\phi}\rangle .
\end{eqnarray}
An evolution of the expectation value of the operator $\hat{w}$ is found to be
\begin{equation}\label{frwda2}
    \langle f (\phi)|\hat{w}(\phi+\psi)f(\phi)\rangle = A \cosh3\kappa(\psi+B),
\end{equation}
where $f:=\sum\alpha_a f_a, \; f_a \in \mathcal{F}_b$. \\  One may verify that
\begin{equation}
    A=\textrm{sgn}(X)\sqrt{X^2-Y^2},~~~~B=\frac{1}{6\kappa}\ln{\frac{X+Y}{X-Y}}~,
\end{equation}
where
\begin{equation}
    X:=\sum_{a}|\alpha_a|^2a,~~~~Y:=\sum_{a,~m}\frac{\bar{\alpha}_b\alpha_a-
    \bar{\alpha}_a\alpha_b}
   {i\pi}\frac{m(2a+6m\kappa^2\gamma\hbar\lambda)}{(2m-1)(2m+1)},
\end{equation}
and where $b = a + 6\kappa^2\gamma\lambda\hbar, \;\; b\in \dR, \;
~m\in \mathbb{Z}$, and $|X|>|Y|$.

One can see that the evolution of the expectation value of the
operator $\hat{w}$ coincides with the classical expression
(\ref{frwvol1}).

\subsection{Summary}

The resolution of the cosmological singularity is due to the loop modification of the Hamiltonian already at the classical level. This modification is parameterized by a continuous parameter $\lambda$, which value is so far unknown. Each value of that parameter specifies quantum of the volume and the maximum energy density of the matter (scalar field in our model).\\ Spectrum of the the volume operator is bounded from below and discrete. Its expectation value coincides with the classical expression.\\ The spectrum of the energy density is bounded from below and continuous. There is a coincidence between classical and quantum expressions for the maximum energy density.

\bigskip

\def\baselinestretch{1}

\chapter{Bianchi I model in terms of nonstandard LQC}

\def\baselinestretch{1.66}


From Introduction we know that the Bianchi I model of the Universe is of primary importance as it underlies, to some extent, the Belinskii-Khalatnikov-Lifshitz (BKL) scenario
\cite{PrzegBKL3,Khalatnikov:2008zt,Montani:2007vu,Garfinkle:2007rv,Erickson:2003zm},
which is believed to describe the Universe in the vicinity of the
cosmological singularity. It has been examined recently within the nonstandard LQC \cite{Dzierzak:2009dj, all}, and has been also studied in the context of the standard LQC \cite{Ashtekar:2009vc,Bojowald:2003md,Chiou:2006qq,Szulc:2008ar,MartinBenito:2008wx}.
Clear exposition of the singularity aspects of the Bianchi I model can be found in \cite{Chiou:2007mg,Erickson:2003zm}.

This chapter presents analyzes of the Bianchi I model within the nonstandard LQC formalism. In what follows we consider the Bianchi I model with free massless scalar field in $T^3$ topology. We choose this topology, because taking $\dR^3$ might lead to problems with interpretation of the spectra of the volume operator \cite{all}.

\smallskip

\goodbreak
\section{Classical level}

\subsection{Hamiltonian}

The Bianchi I model with massless scalar field is described by
the metric:
\begin{equation}\label{bibia}
ds^2= -N^2\,dt^2 + \sum_{i=1}^{3} a_i^2(t)\,dx_i^2,
\end{equation}
where
\begin{equation}\label{bibbb}
a_i(\tau)=
a_i(0)\,\bigg(\frac{\tau}{\tau_0}\bigg)^{\textrm{k}_i},~~~~d\tau=
N\,dt,~~~~\sum_{i=1}^{3} \textrm{k}_i= 1 = \sum_{i=1}^{3}
\textrm{k}_i^2 + \textrm{k}_{\phi}^2,
\end{equation}
and where $\,\textrm{k}_{\phi}$ describes matter field density
($\textrm{k}_{\phi}= 0$ corresponds to the Kasner model).

To make this chapter self-contained we remind that in GR the gravitational part of the classical Hamiltonian, $H_g$, is a linear combination of the first-class
constraints
\begin{equation}\label{bihham1}
    H_g:= \int_\Sigma d^3 x (N^i C_i + N^a C_a + N C),
\end{equation}
where $\Sigma$ is the spacelike part of spacetime $\dR \times \Sigma$, $~(N^i, N^a, N)$ denote Lagrange multipliers, $(C_i, C_a, C)$ are the Gauss, diffeomorphism and scalar constraint functions. In our notation  $(a,b = 1,2,3)$ are spatial and $(i,j,k = 1,2,3)$
are internal $SU(2)$ indices. As we know the constraints must satisfy a
specific algebra.

Having fixed local gauge and diffeomorphism freedom we can rewrite it for the
Bianchi I model with massless scalar field
\cite{Dzierzak:2009dj}
\begin{equation}\label{bihamG}
H_g = - \gamma^{-2} \int_{\mathcal V} d^3 x ~N
e^{-1}\varepsilon_{ijk}
 E^{aj}E^{bk} F^i_{ab}\, ,
\end{equation}
where  $\gamma$ is the Barbero-Immirzi parameter, $\mathcal
V\subset \Sigma$ is an elementary cell, $\Sigma$ is spacelike
hypersurface,  $N$ denotes the lapse function, $\varepsilon_{ijk}$
is the alternating tensor, $E^a_i $ is a densitized vector field,
$e:=\sqrt{|\det E|}$, and where $F^i_{ab}$ is the curvature of an
$SU(2)$ connection $A^i_a$.

As we know, the resolution of the singularity, obtained within LQC, is based
on rewriting the curvature $F^k_{ab}$ in terms of holonomies
around loops. The curvature $F^k_{ab}$ may be determined
by making use of the formula
\begin{equation}\label{bicur}
F^k_{ab}= -2~\lim_{Ar\,\Box_{ij}\,\rightarrow \,0}
Tr\;\Big(\frac{h_{\Box_{ij}}-1}{Ar\,\Box_{ij}}\Big)\;{\tau^k}\;
^o\omega^i_a  \; ^o\omega^j_a ,
\end{equation}
where
\begin{equation}\label{bibox}
h_{\Box_{ij}} = h^{(\mu_i)}_i h^{(\mu_j)}_j (h^{(\mu_i)}_i)^{-1}
(h^{(\mu_j)}_j)^{-1}
\end{equation}
is the holonomy of the gravitational connection around the square
loop $\Box_{ij}$,  considered over a face of the elementary cell,
each of whose sides has length $\mu_j L_j$ (and $V_o:=L_1 L_2
L_3$) with respect to the flat fiducial metric $^o q_{ab}:=
\delta_{ij}\, ^o \omega^i_a\, ^o \omega^j_a $; the fiducial triad
$^o e^a_k$ and cotriad $^o \omega^k_a$ satisfy $^o \omega^i_a\,^o
e^a_j = \delta^i_j$; $~Ar\,\Box_{ij}$ denotes the area of the
square; and $V_o = \int_{\mathcal V} \sqrt{^o q} d^3 x$ is the
fiducial volume of $\mathcal V$.

The holonomy in the fundamental, $j=1/2$, representation of
$SU(2)$ reads
\begin{equation}\label{bihol}
h^{(\mu_i)}_i  =\cos (\mu_i c_i/2)\;\id + 2\,\sin (\mu_i
c_i/2)\;\tau_i,
\end{equation}
where $\tau_i = -i \sigma_i/2\;$ ($\sigma_i$ are the Pauli spin
matrices). The connection $A^k_a$ and the density weighted
triad $E^a_i$ (which occurs in (\ref{biidentT})) are
determined by the conjugate variables $c$ and $p$:
\begin{equation}
A^i_a = \,c_i\,{L_i}^{-1}\,^o\omega^i_a, ~~~~E^a_i = \,p_i\,{L_j}^{-1}
\,{L_k}^{-1}\,^oe^a_i ,
\end{equation}
where:
\begin{equation}\label{bipici}
c_i = \gamma\,\dot{a_i}\,L_i,~~~~ |p_i| = a_j\,a_k\,L_j\,L_k .
\end{equation}
and
\begin{equation}
\{c_i, p_j\}= 8\pi G \gamma \delta_{ij}
\end{equation}

Making  use of (\ref{bihamG}), (\ref{bicur})  and the so-called
Thiemann identity
\begin{equation}\label{biidentT}
\varepsilon_{ijk}\,e^{-1}\,E^{aj}E^{bk} =
\frac{\text{sgn}(p_1p_2p_3)}{2\pi G \gamma
(\mu_1\mu_2\mu_3)^{1/3}}\,\sum_k\,^o\varepsilon^{abc}\,
^o\omega^k_c\,Tr
\Big(h_k^{(\mu_k)}\{(h_k^{(\mu_k)})^{-1},V\}\,\tau_i \Big)
\end{equation}
leads to $H_g$ in the form
\begin{equation}\label{bihamR}
    H_g = \lim_{\mu_1,\mu_2,\mu_3\rightarrow \,0}\; H^{(\mu_1\,\mu_2\,\mu_3)}_g ,
\end{equation}
where
\begin{eqnarray}\label{bihamL}
H^{(\mu_1\,\mu_2\,\mu_3)}_g = - \frac{\text{sgn}(p_1p_2p_3)}{2\pi
G \gamma^3 \mu_1\mu_2\mu_3} \sum_{ijk}\,N\, \varepsilon^{ijk}\, Tr
\Big(h^{(\mu_i)}_i h^{(\mu_j)}_j (h^{(\mu_i)}_i)^{-1}
(h^{(\mu_j)}_j)^{-1} \nonumber\\h_k^{(\mu_k)}\{(h_k^{(\mu_k)})^{-1},V\}\Big),
\end{eqnarray}
and where $V= a_1\,a_2\,a_3\,L_1L_2L_3$ is the volume of the elementary cell
$\mathcal{V}$.

The total Hamiltonian for Bianchi I universe with a massless
scalar field, $\phi$, reads
\begin{equation}\label{biham}
   H = H_g + H_\phi \approx 0,
\end{equation}
where $H_g$ is defined by (\ref{bihamR}). The Hamiltonian of the
scalar field  is known to be: $H_\phi = N\,p^2_\phi
|p_1p_2p_3|^{-\frac{1}{2}}/2$, where $\phi$ and $p_\phi$ are the
elementary variables satisfying $\{\phi,p_\phi\} = 1$. The
relation $H \approx 0$ defines the  physical phase space of
considered gravitational system with constraints.

Making use of (\ref{bihol}) we calculate  (\ref{bihamL}) and get the
modified  total Hamiltonian $H^{(\lambda)}_g$ corresponding
to (\ref{biham}) in the form
\begin{equation}\label{biregH}
  H^{(\lambda)}/N= -\frac{1}{8\pi G \gamma^2}\;\frac{\text{sgn}(p_1p_2p_3)}
  {\mu_1\mu_2\mu_3}\bigg[\sin(c_1 \mu_1)\sin(c_2
\mu_2)\,\mu_3\;\textrm{sgn}(p_3)\sqrt{\frac{|p_1p_2|}{|p_3|}} +
\textrm{cyclic}\bigg] + \frac{p_{\phi}^2}{2\,V}
\end{equation}
where
\begin{equation}\label{bire1}
\mu_i:= \sqrt{\frac{1}{|p_i|}}\,\lambda ,
\end{equation}
and where $\lambda$ is a regularization parameter. Here we
wish to emphasize that (\ref{biregH}) presents a loop modified but classical Hamiltonian.

It is known \cite{Ashtekar:2009vc,Chiou:2007mg,Szulc:2008ar} that such a choice of $\mu_i$ leads to the dependence of the final results on the fiducial volume $V_0$. In the universe with compact topology, like that considered by us, $V_0$ has physical sense, in contrast to a case with noncompact topology. Thus, an expected dependence of the results on $V_0$ would be rather meritorious than problematic.

In the gauge $N= \sqrt{|p_1\,p_2\,p_3|}$ the Hamiltonian modified by loop geometry reads
\begin{equation}
H^{(\lambda)}= -\frac{1}{8\pi G
\gamma^2\lambda^2}\;\bigg[|p_1p_2|^{3/2}\sin(c_1 \mu_1)\sin(c_2
\mu_2) + \textrm{cyclic} \bigg] + \frac{p_{\phi}^2}{2} .
\end{equation}

The Poisson bracket is defined to be
\begin{equation}\label{bire2}
    \{\cdot,\cdot\}:= 8\pi G\gamma\;\sum_{k=1}^3\bigg[ \frac{\partial \cdot}
    {\partial c_k} \frac{\partial \cdot}{\partial p_k} -
     \frac{\partial \cdot}{\partial p_k} \frac{\partial \cdot}{\partial c_k}\bigg] +
     \frac{\partial \cdot}{\partial \phi} \frac{\partial \cdot}{\partial p_\phi} -
     \frac{\partial \cdot}{\partial p_\phi} \frac{\partial \cdot}{\partial
     \phi} ,
\end{equation}
where $(c_1,c_2,c_3,p_1,p_2,p_3,\phi,p_\phi)$ are canonical
variables.  The dynamics of  $\xi$ reads
\begin{equation}\label{bidyn}
    \dot{\xi} := \{\xi,H^{(\lambda)}\},~~~~~~\xi \in
    \{c_1,c_2,c_3,p_1,p_2,p_3,\phi,p_\phi\}.
\end{equation}
The dynamics in the physical phase space,
$\mathcal{F}_{phys}^{(\lambda)}$, is defined by solutions to
(\ref{bidyn}) satisfying the condition $H^{(\lambda)}\approx 0$. The
solutions of (\ref{bidyn}) ignoring the constraint
$H^{(\lambda)}\approx 0$ are in the kinematical phase space,
$\mathcal{F}_{kin}^{(\lambda)}$.

We use the following canonical variables
\begin{equation}\label{biham1}
\beta_i := \frac{c_i}{\sqrt{|p_i|}},~~~~v_i:=
|p_i|^{3/2},
\end{equation}
where $i = 1,2,3$. They satisfy the algebra
\begin{equation}\label{biham2}
\{\beta_i, v_j\}= 12\pi G\gamma\delta_{ij},
\end{equation}
where the Poisson bracket reads
\begin{equation}\label{biham3}
\{\cdot,\cdot\} = 12\pi G\gamma\;\sum_{k=1}^3\bigg[ \frac{\partial
\cdot}{\partial \beta_k} \frac{\partial \cdot}{\partial v_k} -
\frac{\partial \cdot}{\partial v_k} \frac{\partial \cdot}{\partial
\beta_k}\bigg] + \frac{\partial \cdot}{\partial \phi}
\frac{\partial \cdot}{\partial p_\phi} - \frac{\partial
\cdot}{\partial p_\phi} \frac{\partial \cdot}{\partial \phi} .
\end{equation}
The Hamiltonian in the variables (\ref{biham1}) turns out to be
\begin{eqnarray}\nonumber
H^{(\lambda)}= \frac{p_{\phi}^2}{2} - \frac{1}{8\pi
G\gamma^2}\bigg(\frac{\sin(\lambda\beta_1)\sin(\lambda\beta_2)}{\lambda^2}v_1v_2
+\frac{\sin(\lambda\beta_1)\sin(\lambda\beta_3)}{\lambda^2}v_1v_3\\
\label{bihamNN}
+\frac{\sin(\lambda\beta_2)\sin(\lambda\beta_3)}{\lambda^2}v_2v_3
\bigg),
\end{eqnarray}
where $\lambda$ parameterizes the holonomy of connection modifying
the Bianchi I model.

\subsection{Classical dynamics}

The Hamilton equations of motion read
\begin{eqnarray}
\dot{\beta_i}&=&
-18\pi G\,\frac{\sin(\lambda\beta_i)}{\lambda}\,(\textrm{O}_j+\textrm{O}_k),\label{bibeta}\\
\dot{v_i}&=& 18\pi G\,v_i\cos(\lambda\beta_i)\,(\textrm{O}_j+\textrm{O}_k),\label{biv}\\
\dot{\phi}&=& p_{\phi},\label{biphi}\\
\dot{p_{\phi}}&=& 0,\\
\label{bidyn1} H^{(\lambda)}&\approx&0,
\end{eqnarray}
where $i,j,k=1,2,3$ and $i\neq j\neq k$ and where
\begin{equation}\label{birow1}
\textrm{O}_i:= \frac{v_i\,\sin(\lambda\beta_i)}{12\pi G
\gamma\lambda}.
\end{equation}

Insertion of (\ref{biv}) into (\ref{bibeta}) gives
\begin{eqnarray}
d\beta_i= -\frac{\tan(\lambda\beta_i)}{\lambda}\,\frac{dv_i}{v_i},
\end{eqnarray}
which leads to
\begin{equation}\label{birow2}
v_i\,\frac{\sin(\lambda\beta_i)}{\lambda} = \textrm{const}
\end{equation}
Therefore, $\textrm{O}_i$ are constants of motion.

Making use of (\ref{biphi}), (\ref{biv}) and $\;\cos(\lambda\beta_i)=
\sqrt{1-\sin(\lambda\beta_i)^2}\;$ gives
\begin{equation}\label{bieq1}
\int\frac{d v_i}{\sqrt{v_i^2-(12\pi
G\gamma\lambda\,\textrm{O}_i)^2}}=18\pi G
\int\frac{(\textrm{O}_j+\textrm{O}_k)}{p_{\phi}}\,d\phi.
\end{equation}

Integration of  (\ref{bieq1}) leads to
\begin{eqnarray}
\ln\bigg|v_i + \sqrt{v_i^2-(12\pi
G\gamma\lambda\,\textrm{O}_i)^2}\bigg|= \frac{18\pi
G}{p_{\phi}}\,(\textrm{O}_j+\textrm{O}_k)\,(\phi
- \phi^{0}_i).
\end{eqnarray}
Thus we have
\begin{eqnarray}
2\,|v_i|= \exp\bigg(\frac{18\pi
G}{p_{\phi}}\,(\textrm{O}_j+\textrm{O}_k)\,(\phi
- \phi^{0}_i)\bigg) + (12\pi
G\gamma\lambda\textrm{O}_i)^2\times\\\nonumber
\times\exp\bigg(-\frac{18\pi
G}{p_{\phi}}\,(\textrm{O}_j+\textrm{O}_k)\,(\phi
- \phi^{0}_i)\bigg),
\end{eqnarray}
which may be rewritten as
\begin{equation}\label{bieq2}
v_i= 12\pi
G\gamma\lambda\,\,|\textrm{O}_i|\,\cosh\bigg(\frac{18\pi
G}{p_{\phi}}\,(\textrm{O}_j+\textrm{O}_k)\,(\phi
- \phi^{0}_i) - \ln\big|12\pi G\gamma\lambda\textrm{O}_i
\big|\bigg).
\end{equation}

\subsection{Elementary observables}

As we know, function $F$ defined on the phase space is a Dirac observable if
it is a solution to the equation
\begin{equation}\label{biob1}
\left\{F, H^{(\lambda)}\right\} \approx 0.
\end{equation}
An explicit form of (\ref{biob1}) is given by
\begin{eqnarray}
12\pi G\gamma\sum_{i=1}^{3}\bigg(\frac{\partial F}{\partial
\beta_i}\frac{\partial H^{(\lambda)}}{\partial v_i} -
\frac{\partial F}{\partial v_i}\frac{\partial
H^{(\lambda)}}{\partial \beta_i} \bigg) + \frac{\partial
F}{\partial \phi}p_{\phi} = 0,
\end{eqnarray}
which reads
\begin{eqnarray}
18\pi G\,\sum_{i=1}^{3}\bigg[v_i\cos(\lambda\beta_i)\frac{\partial
F}{\partial v_i} -
\frac{\sin(\lambda\beta_i)}{\lambda}\frac{\partial F}{\partial
\beta_i} \bigg]\cdot(\textrm{O}_j+\textrm{O}_k)
+ \frac{\partial F}{\partial \phi}p_{\phi}= 0 \label{birowobs}.
\end{eqnarray}

\subsubsection{Kinematical observables}

One may easily verify that $\textrm{O}_i$ satisfy
(\ref{birowobs}). Instead of solving (\ref{birowobs}) one may use the
constants that occur in (\ref{bieq2}). This way we get
\begin{equation}\label{bikin1}
\textrm{A}_i =
\ln{\bigg|\frac{\tan\big(\frac{\lambda\beta_i}{2}\big)}{\frac{\lambda}{2}}\bigg|}
+18\pi
G\,\frac{(\textrm{O}_j+\textrm{O}_k)}{p_{\phi}}\,\phi
\end{equation}

The observables (\ref{bikin1}) are called kinematical as they
are not required to satisfy the constraint (\ref{bidyn1}).

\subsubsection{Dynamical observables}

An explicit form of the constraint (\ref{bidyn1}) in terms of
$\textrm{O}_i$ is given by
\begin{equation}\label{bidy2}
p_{\phi}\,\,\textrm{sgn}(p_{\phi})= 6\sqrt{\pi
G}\,\sqrt{\textrm{O}_1\textrm{O}_2 +
\textrm{O}_1\textrm{O}_3 +
\textrm{O}_2\textrm{O}_3} .
\end{equation}
It results from (22), (27) and (28) that
$\textrm{O}_1\textrm{O}_2 +
\textrm{O}_1\textrm{O}_3 +
\textrm{O}_2\textrm{O}_3 \geq 0$ so (42) is well
defined. Thus, the dynamical observables,
$\textrm{A}_i^{\textrm{dyn}}$, corresponding to
(\ref{bikin1}) read
\begin{equation}\label{bidy3}
\textrm{A}_i^{\textrm{dyn}}=
\ln{\bigg|\frac{\tan\big(\frac{\lambda\beta_i}{2}\big)}{\frac{\lambda}{2}}\bigg|}
+ \frac{3\sqrt{\pi
G}\,\,\textrm{sgn}(p_{\phi})\big(\textrm{O}_j+\textrm{O}_k\big)\,
\phi}{\sqrt{\textrm{O}_1\textrm{O}_2 +
\textrm{O}_1\textrm{O}_3 +
\textrm{O}_2\textrm{O}_3}} .
\end{equation}

\subsubsection{Algebra of elementary observables}

One may verify that  $\textrm{A}_i^{\textrm{dyn}}$
satisfy the following Lie algebra
\begin{eqnarray}
\{\textrm{O}_i, \textrm{O}_j\}&=& 0, \\
\label{biaa1} \{\textrm{A}_i^{\textrm{dyn}},
\textrm{O}_j\}&=& \delta_{ij},  \\\label{biaa3}
\{\textrm{A}_i^{\textrm{dyn}},\textrm{A}_j^{\textrm{dyn}}\}&=&0.
\end{eqnarray}
In the physical phase space the Poisson bracket is found
to be
\begin{equation}\label{bidyyy}
\{\cdot, \cdot\}_{\textrm{dyn}}:=
\sum_{i=1}^{3}\bigg(\frac{\partial \cdot}{\partial
\textrm{A}_i^{\textrm{dyn}}} \frac{\partial
\cdot}{\partial \textrm{O}_i} - \frac{\partial
\cdot}{\partial \textrm{O}_i} \frac{\partial
\cdot}{\partial \textrm{A}_i^{\textrm{dyn}}} \bigg) ,
\end{equation}
and the algebra reads
\begin{eqnarray}\label{bia0}
\{\textrm{O}_i, \textrm{O}_j\}_{\textrm{dyn}}&=& 0, \\
\label{bia1} \{\textrm{A}_i^{\textrm{dyn}},
\textrm{O}_j\}_{\textrm{dyn}}&=& \delta_{ij},
 \\\label{bia3}
\{\textrm{A}_i^{\textrm{dyn}},\textrm{A}_j
^{\textrm{dyn}}\}_{\textrm{dyn}}&=&0.
\end{eqnarray}

\subsection{Compound observables}

In what follows we consider the physical observables which
characterize the singularity aspects of the Bianchi I model. It is
helpful to rewrite (\ref{bidy2}) and (\ref{bieq2}) in the form
\begin{equation}\label{bicom1}
p_{\phi}^2= 36\pi G\,\big( \textrm{O}_1\textrm{O}_2 +
\textrm{O}_1\textrm{O}_3 + \textrm{O}_2\textrm{O}_3 \big),
\end{equation}
\begin{equation}\label{bicom2}
v_i= 12\pi
G\gamma\lambda\,|\,\textrm{O}_i|\,\cosh\bigg(\frac{3\sqrt{\pi
G}\,\,\textrm{sgn}(p_{\phi})\big(\textrm{O}_j+\textrm{O}_k
\big)\,\phi}{\sqrt{\textrm{O}_1\textrm{O}_2 +
\textrm{O}_1\textrm{O}_3 +
\textrm{O}_2\textrm{O}_3}} +
\ln\bigg(\frac{\lambda}{2}\bigg)-
\textrm{A}_i^{\textrm{dyn}}\bigg).
\end{equation}
The so-called directional energy density \cite{Chiou:2007mg} is
defined to be
\begin{equation}\label{bicom3}
\rho_{i}(\lambda,\phi):= \frac{p_{\phi}^2}{2\,v_{i}^2}.
\end{equation}

The bounce in the $i$-th direction occurs when $\rho_{i}$
approaches its maximum \cite{Chiou:2007mg}, which happens at the
minimum of $v_i$ ($p_{\phi}$ is a constant of motion). One may
easily verify that in the case when all three directions coincide,
which corresponds to the FRW model, these densities turn into the energy density of the flat FRW with massless scalar field \cite{Dzierzak:2009ip}.

It is clear that $v_i$ takes minimum  for $\cosh(\cdot)= 1$ so we have
\begin{equation}\label{biden}
v_i^{min}= 12\pi
G\gamma\lambda\,\textrm{O}_i,~~~~\rho_{i}^{max}=
\frac{1}{2} \Big(\frac{p_{\phi}}{12\pi
G\gamma\lambda\,\textrm{O}_i}\Big)^2.
\end{equation}
Rewriting $\textrm{O}_i$ and $p_{\phi}$ in terms of $\textrm{k}_i$ and
$\textrm{k}_\phi$ \cite{Chiou:2007mg}
\begin{equation}\label{bidd}
\textrm{O}_i = \frac{2}{3}\,\textrm{k}_i\,K,~~~~
\textrm{p}_{\phi} = \sqrt{8\pi G}\,\textrm{k}_{\phi}K,
\end{equation}
where $K$ is a constant, leads to
\begin{equation}\label{bidenn}
\rho_{i}^{max}= \frac{1}{16\pi
G\gamma^2\lambda^2}\,\bigg(\frac{\textrm{k}_{\phi}}
{\textrm{k}_i}\bigg)^2 .
\end{equation}
We can determine $\rho_{i}^{max}$ if we know $\lambda$, but as we remember $\lambda$ is a free parameter of the formalism.

One may apply (\ref{bidenn}) to the Planck scale. Substituting
$\lambda=l_{Pl}$  gives
\begin{equation}\label{bide}
\rho_{i}^{max}\simeq
0,35\,\bigg(\frac{\textrm{k}_{\phi}}{\textrm{k}_i}\bigg)^2\,\rho_{Pl},
\end{equation}
which demonstrates that $\rho_{i}^{max}$ may fit the Planck scale depending on the ratio $\textrm{k}_{\phi}/\textrm{k}_i$.\\

Another important physical  observable is the volume of the
Universe. From the definitions (\ref{bipici}) and (\ref{biham1}) we
get
\begin{equation}
V= a_1 a_2 a_3 = (v_1v_2v_3)^{1/3} .
\end{equation}
It is clear from (\ref{bicom2}), (\ref{bidd}) and (\ref{bibbb}) that the
volume is bounded from below.

\bigskip

\goodbreak
\section{Preparations to quantization}

This section is devoted to some redefinitions which are helpfull to the procedure of a quantization. The quantization is required despite the fact that the singularity problem is resolved already at the classical level due to the modifications based on the loop geometry. The reasons are described above. Here we analyze the structure of the phase space, which is obviously much more complicated than in the FRW case.

\subsection{Redefinitions}

\subsubsection{Redefinition of evolution parameter}

Firstly we slightly redefine the elementary Bianchi observables used in previous sections. Now they read
\begin{equation}\label{biO2}
\textrm{O}_{i}:=
\frac{1}{3\kappa\gamma}\,\frac{v_i\sin(\lambda\beta_i)}{\lambda},
\end{equation}
and
\begin{equation}\label{biA2}
\textrm{A}_i:=
\frac{1}{3\kappa}\ln{\bigg(\frac{\big|\tan\big(\frac{\lambda\beta_i}{2}
\big)\big|}{\frac{\lambda}{2}}\bigg)} + \frac{3}{2\sqrt{3}}
\frac{\textrm{sgn}(p_{\phi})\big(\textrm{O}_j+\textrm{O}_k\big)\,
\phi}{\sqrt{\textrm{O}_1\textrm{O}_2 + \textrm{O}_1\textrm{O}_3 +
\textrm{O}_2\textrm{O}_3}},
\end{equation}
where $\kappa^2 := 4\pi G/3$. One may verify that the algebra of redefined observables is isomorphic to the previous one
\begin{equation}\label{bialg prim}
\{\textrm{O}_{i},\textrm{O}_{j}\}=
0,~~~~\{\textrm{A}_{i},\textrm{O}_{j}\}=
\delta_{ij},~~~~\{\textrm{A}_{i},\textrm{A}_{j}\} = 0 .
\end{equation}
and
\begin{equation}\label{bivi1}
v_i= 3\kappa\gamma\lambda|\textrm{O}_{i}|\,\cosh\bigg(\frac{3\sqrt{\pi
G}\,\big(\textrm{O}_{j}+\textrm{O}_{k}
\big)\,\phi}{\sqrt{\textrm{O}_{1}\textrm{O}_{2} +
\textrm{O}_{1}\textrm{O}_{3} + \textrm{O}_{2}\textrm{O}_{3}}} -
3\kappa\textrm{A}_{i}\bigg) .
\end{equation}

Since the observables $\textrm{O}_i$ are constants of
motion in $\phi \in \dR$, it is possible to make the
following redefinition of an evolution parameter
\begin{equation}\label{biphip}
\varphi:= \frac{\sqrt{3
}\,\,\phi}{2\,\sqrt{\textrm{O}_\textrm{1}\textrm{O}_\textrm{2} +
\textrm{O}_\textrm{1}\textrm{O}_\textrm{2} +
\textrm{O}_\textrm{2}\textrm{O}_\textrm{3}}}
\end{equation}
so we have
\begin{equation}\label{bivi2}
v_i= 3\kappa\gamma\lambda|\textrm{O}_i|\,\cosh
3\kappa\big((\textrm{O}_j+\textrm{O}_k )\,\varphi -
\textrm{A}_i\big),
\end{equation}
which simplifies further considerations.

\subsubsection{New elementary observables}

One can make the following redefinitions
\begin{equation}\label{biredef}
\mathcal{A}_i := \textrm{A}_i - (\textrm{O}_j +
\textrm{O}_k)\,\varphi.
\end{equation}
Thus, the directional volume  (\ref{bivi2})  becomes
\begin{equation}\label{bivi5}
v_i := |w_i|,~~~~~w_i= 3\kappa\gamma\lambda \textrm{O}_i\,\cosh (3 \kappa
\mathcal{A}_i).
\end{equation}

The algebra of observables reads
\begin{equation}\label{bialc}
\{\textrm{O}_i,\textrm{O}_j\}=
0,~~~~\{\mathcal{A}_i,\textrm{O}_j\}=
\delta_{ij},~~~~\{\mathcal{A}_i,\mathcal{A}_j\}= 0 ,
\end{equation}
where the Poisson bracket  is defined to be
\begin{equation}\label{bipo}
\{\cdot,\cdot\}:= \sum_{k=1}^3\Big(\frac{\partial\cdot}{\partial
\mathcal{A}_k} \frac{\partial\cdot}{\partial \textrm{O}_k} -
\frac{\partial\cdot}{\partial \textrm{O}_k} \frac{\partial\cdot}{\partial
\mathcal{A}_k}\Big).
\end{equation}

\subsection{Structure of phase space}

All considerations carried out in the previous section have been
done under the assumption that the observables $\textrm{O}_1$,
$\textrm{O}_2$ and $\textrm{O}_3$ have no restrictions.  The
inspection of (\ref{biA2}), (\ref{bivi1}) and (\ref{biredef}) shows that
the domain of definition of the elementary observables reads
\begin{equation}\label{bidom}
    D := \{(\mathcal{A}_k, \textrm{O}_k)\,|\,\mathcal{A}_k \in
    \dR,~~ \textrm{O}_1 \textrm{O}_2 + \textrm{O}_1\textrm{O}_3 +
    \textrm{O}_2 \textrm{O}_3 > 0\},
\end{equation}
where $k=1,2,3$. The restriction $\textrm{O}_1 \textrm{O}_2 +
\textrm{O}_1\textrm{O}_3 + \textrm{O}_2 \textrm{O}_3 > 0$ is a
consequence of the Hamiltonian constraint (see,
\cite{Dzierzak:2009dj} for more details).

In what follows we consider two cases:
\begin{enumerate}
    \item Kasner-unlike dynamics: (a) $\textrm{O}_i>0$, $\textrm{O}_j>0$,
    $\textrm{O}_k>0$,
    which describes all three directions expanding (b) $\textrm{O}_i<0$,
    $\textrm{O}_j<0$,
    $\textrm{O}_k<0$, with all directions shrinking.
    \item Kasner-like dynamics: (a) $\textrm{O}_i>0$, $\textrm{O}_j>0$,
    $\textrm{O}_k<0$, which describes two directions expanding and one
    direction shrinking; (b) $\textrm{O}_i <0$, $\textrm{O}_j <0$,
    $\textrm{O}_k >0$, with two directions shrinking and one
    expanding.
\end{enumerate}
This classification presents all possible nontrivial cases. Our
terminology fits the one used in \cite{Chiou:2007mg} due to the
relation $\,\textrm{O}_i= 6 \kappa \textrm{k}_i K,~~(0<K=const)$, where
constants $\textrm{k}_i$ are defined by (\ref{bibbb}).

For more details see Appendix B.

\subsection{True Hamiltonian}

Now we can define a generator of an evolution called a true
Hamiltonian $\mathbb{H}$. Making use of (\ref{biredef}), and
$\textrm{O}_i = const$ (see \cite{Dzierzak:2009dj}), we get
\begin{equation}\label{bitrue}
\{\mathcal{A}_i,\mathbb{H}\}:=\frac{d\mathcal{A}_i}{d\varphi} =-
(\textrm{O}_j + \textrm{O}_k),~~~~
\{\textrm{O}_i,\mathbb{H}\}:=\frac{d\textrm{O}_i}{d\varphi} =0 .
\end{equation}
The solution to (\ref{bitrue}) is easily found to be
\begin{equation}\label{biHtrue}
\mathbb{H}=\textrm{O}_\textrm{1}\textrm{O}_\textrm{2} +
\textrm{O}_\textrm{1}\textrm{O}_\textrm{3} +
\textrm{O}_\textrm{2}\textrm{O}_\textrm{3}.
\end{equation}
The true Hamiltonian is defined on the reduced phase space
which is devoid of constraints. It generates a flow in the
family of volume quantities, enumerated by the evolution
parameter.

\bigskip

\goodbreak
\section{Quantum level}

\subsection{Representation of elementary observables}

We use the Schr\"{o}dinger representation for the algebra
(\ref{bialc}) defined as
\begin{equation}\label{biosy}
\textrm{O}_k\rightarrow\widehat{\textrm{O}}_k\,f_k(x_k):=
\frac{\hbar}{i}\,\frac{d}{dx_k}\,f_k(x_k),~~~\mathcal{A}_k\rightarrow
\widehat{\mathcal{A}}_k\,f_k(x_k):=x_k\, f_k(x_k),~~~k=1,2,3.
\end{equation}

One may verify that
\begin{equation}\label{bialmod}
[\widehat{\textrm{O}}_i,\widehat{\textrm{O}}_j]=
0,~~~~[\widehat{\mathcal{A}}_i,\widehat{\mathcal{A}}_j]=0,~~~~
[\widehat{\mathcal{A}}_i,\widehat{\textrm{O}}_j]=
i\hbar \,\delta_{ij}.
\end{equation}
The representation is defined formally on some dense subspaces of
a Hilbert space to be specified later.

\subsection{Kasner-unlike case}

The condition $\textrm{O}_1 \textrm{O}_2 +
\textrm{O}_1\textrm{O}_3 + \textrm{O}_2 \textrm{O}_3 > 0$ is
automatically  satisfied in this case, because $\textrm{O}_1,
\textrm{O}_2$ and $\textrm{O}_3$ are of the same sign. To be
specific, let us consider (1a); the case (1b) can be done by
analogy.

Let us quantize the directional volumes by means of $w_i$ defined
in (\ref{bivi5}). A standard procedure gives\footnote{Subscripts of observables are dropped to simplify notation.}
\begin{eqnarray}\label{biqdir}
\hat{w}&:=&
    \frac{3\kappa\gamma\lambda}{2}\,
\bigg(\widehat{\textrm{O}}\,\cosh \big(3\kappa\widehat{\mathcal{A}}\big)
    + \cosh\big(3\kappa\widehat{\mathcal{A}}\big)\;\widehat{\textrm{O}}\bigg)=\nonumber\\
    &=& -\frac{ia}{2}\Big(2\cosh(bx)\frac{d}{dx}+b\sinh(bx)\Big),
\end{eqnarray}
where  $a:=3\kappa\gamma\lambda\hbar\,$ and $b:=3\kappa$, and where
we have used the representation for the elementary observables
defined by (\ref{biosy}).

In what follows we solve the eigenvalue problem for the operator
$\hat{w}$ and identify its domain of self-adjointness.

Let us consider the invertible mapping $L^2(\mathbb{R},dx)\ni \psi
\rightarrow \tilde{U}\psi=: f \in L^2(\mathbb{I},dy)$ defined by
\begin{equation}\label{biinterval}
 \tilde{U}\psi (x) :=\frac{\psi(\ln|\textrm{tg}^{1/b}(\frac{by}{2})|)}{\sin^{1/2}(by)}
 =:  f(y),~~x\in
    \mathbb{R},~~y\in \mathbb{I} := (0,\pi/b).
\end{equation}
We have
\begin{eqnarray}
\nonumber
    \langle\psi|\psi\rangle=\int_{-\infty}^{\infty}\overline{\psi}\psi
    ~dx\\ \nonumber
    =\int_0^{\frac{\pi}{b}}\overline{\psi}(\ln|\textrm{tg}^{1/b}(\frac{by}
    {2})|)\psi(\ln|\textrm{tg}^{1/b}(\frac{by}{2})|)d\big(\ln|\textrm{tg}^{1/b}
    (\frac{by}{2})|)\big)\\
    \nonumber
    =\int_0^{\frac{\pi}{b}}\overline{\psi}(\ln|\textrm{tg}^{1/b}(\frac{by}{2})|)
    \psi(\ln|\textrm{tg}^{1/b}(\frac{by}{2})|)\frac{dy}{\sin(by)}\\ \label{scalarproduct}
=\int_0^{\frac{\pi}{b}}\overline{\frac{\psi(\ln|\textrm{tg}^{1/b}(\frac{by}
{2})|)}{\sin^{1/2}(by)}}\frac{\psi(\ln|\textrm{tg}^{1/b}(\frac{by}{2})|)}
{\sin^{1/2}(by)}~dy= \langle \tilde{U}\psi|\tilde{U}\psi\rangle .
\end{eqnarray}
Thus, the mapping (\ref{biinterval}) is isometric and hence unitary.

Now, let us see how the operator $\hat{w}$ transforms under the
unitary map (\ref{biinterval}). The transformation consists of the
change of an independent variable
\begin{equation}\label{biindep}
    x\mapsto y:=\frac{2}{b}\textrm{arctan}(e^{bx}),
\end{equation}
which leads to
\begin{equation}\label{biNEWoperator1}
   -\frac{ia}{2}\Big(2\cosh(bx)\frac{d}{dx}+b\sinh(bx)\Big)\mapsto
   -ia\frac{d}{dy}+i\frac{ab}{2}\cot(by),
\end{equation}
and re-scaling with respect to a dependent variable
\begin{equation}\label{biNEWoperator2}
    -ia\frac{d}{dy}+i\frac{ab}{2}\cot(by)\mapsto\sin^{-1/2}(by)
\bigg(-ia\frac{d}{dy}+i\frac{ab}{2}\cot(by)\bigg)\sin^{1/2}(by)
=-ia\frac{d}{dy}.
\end{equation}
In the process of mapping
\begin{equation}\label{bitran}
\hat{w}\mapsto \tilde{U}\,\hat{w}\,\tilde{U}^{-1} =
-ia\frac{d}{dy} =: \breve{w},
\end{equation}
we have used two identities: $\sin(by) = 1/\cosh(bx)$ and
$\sinh(bx)= - \cot(by)$.

Since $w>0$ (for $\textrm{O}>0$), we assume that the
spectrum of $\breve{w}$ consists of positive eigenvalues. To
implement this assumption, we  define
$\breve{w}:=\sqrt{\breve{w}^2}$ and consider the eigenvalue
problem
\begin{equation}\label{biw^2}
- a^2\frac{d^2}{dy^2}f_{\nu}=\nu^2 f_{\nu},~~~~y\in (0,\pi/b).
\end{equation}
There are two independent solutions for each value of $\nu^2$
(where $\nu\in \dR$), namely: $\sin(\frac{\nu}{a}y)$ and
$\cos(\frac{\nu}{a}y)$. Removing this degeneracy leads to required
positive eigenvalues of $\breve{w}$. We achieve that in a standard
way by requiring that the eigenvectors  vanish at the boundaries,
i.e, at $y =0$ and $y= \pi/b$. As the result we get the following
spectrum
\begin{equation}\label{bist}
f_{\nu}=N\sin(\frac{\nu}{a}y),~~~~\nu^2=(nab)^2,~~ n= 0, 1,2,\dots
\end{equation}
It should be noted that for $n=0$, the eigenvector is a null state
and thus the lowest eigenvalue is $\nu^2=(ab)^2$. Next, we define
the Hilbert space to be the closure of the span of the
eigenvectors  (\ref{bist}). The operator $\breve{w}^2 = -
a^2\frac{d^2}{dy^2}$ is essentially self-adjoint on this span by
the construction. Due to the spectral theorem \cite{RS} we may
define an essentially self-adjoint operator $\breve{w}=\sqrt{-
a^2\frac{d^2}{dy^2}}$ as follows
\begin{equation}\label{bispec}
\breve{w}f_{\nu}:=\nu f_{\nu},~~~~\nu =ab,\,2ab,\,3ab,\dots
\end{equation}

We have considered the case  $w>0$. The case $w<0$ does not
require changing of the Hilbert space. The replacement
$\hat{w}\mapsto -\hat{w}$ leads to  $\nu \mapsto -\nu$.

Finally, we find that the inverse mapping from $L^2(\id,dy)$ to
$L^2(\dR,dx)$ for the eigenvectors of $\breve{w}$ yields
\begin{equation}\label{biinvmap}
    \sin\big(\frac{\nu}{a}y\big)=f_{\nu}(y)\mapsto
     \tilde{U}^{-1}f_{\nu}(y):={\psi}_{\nu}(x)=\frac{\sin\big(\frac{2\nu}
    {ab}\textrm{arctg}
    (e^{bx})\big)}{\cosh^{1/2}(bx)}.
\end{equation}

\subsection{Kasner-like case}

In the case (2a), the conditions $\textrm{O}_1 \textrm{O}_2 +
\textrm{O}_1\textrm{O}_3 + \textrm{O}_2 \textrm{O}_3 > 0$ with
$\textrm{O}_1<0, \textrm{O}_2
>0, \textrm{O}_3>0$ are satisfied in the following domains\footnote{The case (2b)
can be done by analogy.} for $\textrm{O}_k$
\begin{equation}\label{bidom1}
    \textrm{O}_1 \in\, (-d_1,0),~~~\textrm{O}_2 \in\, (d_2,\infty),~~~\textrm{O}_3 \in\,
    (d_3,\infty),
\end{equation}
where $d_2 > d_1,$ and where $d_3 = d_1 d_2/ (d_2 - d_1)$ so $d_3
> d_1$. The full phase space sector of the Kasner-like evolution
is defined as the union
\begin{equation}\label{bifull}
\bigcup_{0<d_1<d_2}(-d_1,0)\times(d_2,\infty)\times(d_3,\infty)
\end{equation}

In the case of $\textrm{O}_2$ and $\textrm{O}_3$, the restrictions
for domains (\ref{bidom1}) translate into the restrictions for the
corresponding domains for the observables $w_2$ and $w_3$, due to
(\ref{bivi5}), and read
\begin{equation}\label{bidom12}
    w_2 \in\, (D_2,\infty),~~~w_3 \in\,
    (D_3,\infty),
\end{equation}
where $D_2 = \kappa \gamma \lambda d_2$ and $D_3 = \kappa \gamma
\lambda d_3$. Thus, quantization of the $w_2$ and $w_3$
observables can be done by analogy to the Kasner-unlike case. The
spectra of the operators $\hat{w}_2$ and $\hat{w}_3$ are almost
the same as the spectrum defined by (\ref{bispec}) with the only
difference that now $\nu> D_2$ and $\nu> D_3$,
respectively\footnote{Spectra are insensitive to unitary
transformations.}.

The case of $w_1$ requires special treatment.  Let us redefine the
elementary observables corresponding to the 1-st direction as
follows
\begin{equation}\label{bired1}
    \Omega_1 := - \frac{O_1}{b \cosh(b\mathcal{A}_1)},~~~~\Omega_2:=
    \sinh(b\mathcal{A}_1).
\end{equation}
The transformation (\ref{bired1}) is canonical, since
$\{\Omega_1,\Omega_2\}=1$, and invertible. The domains transform
as follows
\begin{equation}\label{bir2}
O_1 \in (-d_1,0),~~~\mathcal{A}_1 \in \dR
~~~~~\longrightarrow~~~~~\Omega_1 \in (0,d_1/b)=:
(0,D_1),~~~\Omega_2 \in \dR.
\end{equation}
The observable $v_1$ in terms of redefined  observables reads
\begin{equation}\label{bir3}
v_1 = \frac{ab}{\hbar}\,\Omega_1\,(1 + \Omega_2^2),~~~~v_1 \in
(0,\infty),
\end{equation}
where $ab/\hbar = 12\pi G\gamma\lambda$. To quantize observables
$\Omega_1$ and $\Omega_2$ we use the Schr\"{o}dinger
representation
\begin{equation}\label{bir4}
\Omega_2 \rightarrow \hat{\Omega}_2f(x):=  -i\hbar\partial_x
f(x),~~~~~\Omega_1 \rightarrow \hat{\Omega}_1f(x):= x f(x),~~~~~f
\in L^2(0,D_1).
\end{equation}

Let us find an explicit form for the operator
$\,\frac{ab}{\hbar}(\widehat{\Omega}_1+\widehat{\Omega_1\Omega_2^2})$,
corresponding to  (\ref{bir3}). Since $\Omega_1>0$, the following
classical equality holds
\begin{equation}\label{binov}
\Omega_1\Omega_2^2=\Omega_1^k\cdot\Omega_2\cdot\Omega_1^{1-k-m}\cdot
\Omega_2\cdot\Omega_1^{m},
\end{equation}
where $m, k\in \dR$. This may lead to many operator orderings at
the quantum level. This issue is further discussed in the
appendix.

We propose the following mapping  (we set $\hbar=1$)
\begin{eqnarray}\label{binovum}
\Omega_1\Omega_2^2 \rightarrow
\widehat{\Omega_1\Omega_2^2}&:=&\frac{1}{2}\bigg( \hat{\Omega}_1^k
\hat{\Omega}_2\,\hat{\Omega}_1^{1-k-m}\hat{\Omega}_2\,\hat{\Omega}_1^m
+ \hat{\Omega}_1^m\hat{\Omega}_2\,\hat{\Omega}_1^{1-k-m}
\hat{\Omega}_2\, \hat{\Omega}_1^k\bigg) =\nonumber\\ &=&
-x\partial^2_{xx}-\partial_x+mkx^{-1},
\end{eqnarray}
which formally ensures the symmetricity of
$\widehat{\Omega_1\Omega_2^2}$. The second equality in
(\ref{binovum}) may be verified via direct calculations.

Now, we define the following unitary transformation $W$
\begin{equation}
L^2([0,D_1],dx)\ni f(x)\mapsto W f(x):=
\sqrt{\frac{y}{2}}f\bigg(\frac{y^2}{4}\bigg)\in
L^2([0,2\sqrt{D_1}],dy).
\end{equation}
One may verify that we have
\begin{eqnarray}
W\partial_x W^{\dag}=\frac{2}{y}\partial_y-\frac{1}{y^2},~~~~
W\partial^2_{xx}
W^{\dag}=\frac{4}{y^2}\partial^2_{yy}-\frac{8}{y^2}
\partial_y+\frac{5}{y^4}~.
\end{eqnarray}

Thus, the operator $W$ transforms (\ref{binovum}) into
\begin{equation}
-\partial^2_{yy}+\frac{1}{y^2}\bigg(4mk-\frac{1}{4}\bigg).
\end{equation}

The eigenvalue problem for
$\widehat{\Omega}_1+\widehat{\Omega_1\Omega_2^2}$ reads
\begin{equation}\label{binov1}
\bigg(-\partial^2_{yy}+\frac{1}{y^2}\bigg(4mk-\frac{1}{4}\bigg)+
\frac{y^2}{4}\bigg)\Phi = \nu \,\Phi.
\end{equation}
Now, we can  see an  advantage of  the chosen ordering
prescription (\ref{binovum}). It enables finding a very simple form
of the volume operator. Taking $k=m= 1/4$ turns (\ref{binov1}) into
\begin{equation}\label{biw2}
\bigg(-\partial^2_{yy}+\frac{y^2}{4}-\nu\bigg)\Phi=0 .
\end{equation}
The problem is mathematically equivalent to the one dimensional
harmonic oscillator in a ``box'' with an edge equal to
$2\sqrt{D_1}$. There are two independent solutions for a given
$\nu$
\begin{eqnarray}
\Phi_{\nu,1}= N_1 e^{-y^2/{4}}~_1F_1\bigg(-\frac{1}{2}\nu+
\frac{1}{4},\frac{1}{2},\frac{y^2}{2}\bigg),\\
\Phi_{\nu,2}= N_2 ye^{-y^2/{4}}
~_1F_1\bigg(-\frac{1}{2}\nu+\frac{3}{4},\frac{3}{2},\frac{y^2}{2}\bigg),
\end{eqnarray}
where $\,_1F_1$ is a hypergeometric confluent function,
$\Phi_{\nu,1}$ and $\Phi_{\nu,2}$ are even and odd cylindrical
functions, respectively. A standard condition for the symmetricity
of the operator  defining the eigenvalue problem (\ref{biw2}) leads
to the vanishing of  the wave functions at the boundaries (as the
box defines the entire size of the $1$-st direction). The solution
(after retrieving of $\hbar$ and $ab$) reads\footnote{We ignore
the solution $\Phi_{\nu,1}$ because it cannot vanish at $y=0$.}.
\begin{equation}\label{bi2w2}
\Phi=N ye^{-\frac{y^2}{4\hbar}}
~_1F_1\bigg(-\frac{1}{2}\frac{\nu}{ab}+\frac{3}{4},\frac{3}{2},
\frac{y^2}{2\hbar}\bigg).
\end{equation}
The solution (\ref{bi2w2}) vanishes at $y=0$ as $\Phi$  is an odd
function. The requirement of vanishing at $y=2\sqrt{D_1}$ leads to
the equation
\begin{equation}\label{biw22}
    _1F_1\bigg(-\frac{1}{2}\frac{\nu}{ab\hbar}+\frac{3}{4},\frac{3}{2},
    \frac{2D_1}{\hbar}\bigg)=0.
\end{equation}
An explicit form of (\ref{biw22}) reads
\begin{equation}\label{bicond}
\sum_{n=0}^{\infty}\frac{\big(-\frac{1}{2}\frac{\nu}{ab}+\frac{3}{4}
\big)_n}{\big(\frac{3}{2}\big)_n}
\bigg(\frac{2D_1}{\hbar}\bigg)^n=0 ,
\end{equation}
where $(a)_n=a(a+1)\dots (a+n-1)$. It results from (\ref{bicond})
that the eigenvalues must satisfy the condition: $\nu\geq ab$.

\subsection{Volume operator}

Classically we have
\begin{equation}\label{bivol 2}
V = |w_1 w_2 w_3|^{1/3} .
\end{equation}
One may verify that $v_k$ Poisson commute and $\hat{v}_k$ commute,
so we can take
\begin{equation}\label{bivol3}
\widehat{V}^3 := \hat{v}_1 \hat{v}_2 \hat{v}_3 = |\hat{w}_1
\hat{w}_2 \hat{w}_3|.
\end{equation}
The eigenfunctions of the operator $\hat{w}_1 \hat{w}_2 \hat{w}_3$
have the form\\
$F^{\lambda_1,\lambda_2,\lambda_3}:={f_1}^{\lambda_1}(x_1){f_2}^{\lambda_2}(x_2)
{f_3}^{\lambda_3}(x_3)$, where ${f_i}^{\lambda_i}(x_i)$ is an
eigenvector of $\hat{w}_i$ with eigenvalue $\lambda_i$. The
closure of the span of $F^{\lambda_1,\lambda_2,\lambda_3}$ is a
Hilbert space, in which $\widehat{V}^3$ is a self-adjoint operator
(by construction).

Due to the spectral theorem on self-adjoint operators \cite{RS},
we have
\begin{equation}\label{bispec1}
 V = (V^3)^{1/3}~~~\longrightarrow~~~\widehat{V} F^{\lambda_1,\lambda_2,\lambda_3}
 := \square\,
 F^{\lambda_1,\lambda_2,\lambda_3},
\end{equation}
where
\begin{equation}\label{bispec2}
 \square:= |\lambda_1 \lambda_2 \lambda_3|^{1/3}.
 \end{equation}

\subsubsection{Kasner-unlike case}

In the Kasner-unlike case we use the formula (\ref{bispec}) to get
 \begin{equation}\label{bispec3}
  \square =  |n_1n_2n_3|^{1/3}\,ab,~~~~~n_1, n_2, n_3 \in
  1,2,3,\dots ,
 \end{equation}
which shows that the spectrum of the volume operator does not have
equally distant levels.  The volume $\square$ equal to zero
is not in the spectrum. There exist a quantum of the volume
which equals $\triangle:=ab = 12\pi G\gamma\lambda\hbar$, and which
defines the lowest value in the spectrum.

\subsubsection{Kasner-like case}

The spectrum in this case reads
\begin{equation}
  \square :=  \bigcup_{0<d_1<d_2}\square_{d_1,d_2},~~~~\square_{d_1,d_2}:= \{\lambda_{d_1}
     \lambda_{d_2}
    \lambda_{d_3}\,|\, d_3 = d_1 d_2/ (d_2 - d_1)\},
\end{equation}
where $\lambda_{d_1}$ is any value subject to the condition
(\ref{bicond}), $\lambda_{d_2}>D_2$ and $\lambda_{d_3}>D_3$ are
given by (\ref{bispec}). The volume $\square$ equal to zero is
not in the spectrum.

\subsection{Evolution}

In this section we ignore the restrictions concerning the domains
of $\textrm{O}_1$, $\textrm{O}_2$ and $\textrm{O}_3$, and we
assume that the Hilbert space of the system is
$L^2(\mathbb{R}^3,dxdydz)$. An  inclusion of the restrictions
would complicate the calculations without bringing any qualitative
change into the picture of evolution.

The generator of evolution determined in (\ref{biHtrue}) may be
formally quantized, due to (\ref{biosy}), as follows
\begin{equation}
\mathbb{H}\mapsto \hat{\mathbb{H}}
=-\hbar^2(\partial_y\partial_z+\partial_x\partial_z+\partial_x\partial_y).
\end{equation}
Since it is self-adjoint in $L^2(\mathbb{R}^3,dxdydz)$, a quantum
evolution can be defined by an unitary operator
\begin{equation}
 U= e^{-i\hbar\tau(\partial_y\partial_z+\partial_x\partial_z+\partial_x\partial_y)},
 ~~~~~\tau \in \dR.
\end{equation}
Let us study  an evolution of the expectation value of the
directional volume $\hat{v}_1$
\begin{equation}
 \langle\psi |U^{-1}\hat{v}_1U | \psi\rangle
\end{equation}
Since $\hat{v}_1$ does not depend on $y$ and $z$, we simplify our
considerations by taking
\begin{equation}
 U_1= e^{-i\hbar\tau(\partial_z+\partial_y)\partial_x}.
\end{equation}
If we are interested in the action of $U_1$ on the functions
$f(x)\in L^2(\mathbb{R},dx)$, then the derivatives
$-i\frac{d}{dy}$ and $-i\frac{d}{dz}$ occurring in $U_1$ commute
and, being self-adjoint, lead finally to real numbers. Let us call
them $k_y$ and $k_z$, respectively, and let us introduce the
parameter $k=k_y+k_z$. Hence, $U_1$ further simplifies and reads
\begin{equation}
 U_1= e^{k\hbar\tau\partial_x}.
\end{equation}
The action of $U_1$ on $f(x)$ reads
\begin{equation}
 U_1f(x)=f(x+k\hbar\tau) .
\end{equation}
We recall that under the unitary mapping
$L^2(\mathbb{R},dx)\mapsto L^2(\mathbb{I},dy)$, defined by
(\ref{biinterval}), the operator $\hat{v}_1$ becomes
$-ia\frac{d}{dy}$ on  $L^2(\mathbb{I},dy)$. Now, let us study an
action of operator $U_1$ on the functions $\varphi(y)\in
L^2(\mathbb{I},dy)$. Straightforward calculation leads to
\begin{equation}
L^2(\mathbb{I},y)\ni\varphi(y)\mapsto\frac{\varphi(\frac{2}{b}\arctan(e^{bx}))}
{\cosh^{1/2}(bx)}\in L^2(\mathbb{R},x),
\end{equation}
and we have
\begin{equation}
 U_1\frac{\varphi(\frac{2}{b}\arctan(e^{bx}))}{\cosh^{1/2}(bx)}=
 \frac{\varphi(\frac{2}{b}\arctan(e^{bx+bk\hbar\tau}))}{\cosh^{1/2}(bx+bk\hbar\tau)}
\end{equation}
The transformation $\tilde{U}^{-1}$ gives
\begin{equation}\label{bievol}
 \frac{\varphi(\frac{2}{b}\arctan(e^{bx+bk\hbar\tau}))}{\cosh^{1/2}(bx+bk\hbar\tau)}
 \mapsto\frac{\varphi(\frac{2}{b}\arctan(e^{bk\hbar\tau}\tan(\frac{by}{2})))}
 {\sqrt{\frac{1}{2}\sin(by)(\tan(\frac{by}{2})e^{bk\hbar\tau}+\cot(\frac{by}{2})
 e^{-bk\hbar\tau})}}=:\varphi_\tau(y),
\end{equation}
where $\varphi_{\tau=0}(y)=\varphi(y)$. Now, we observe that the
symmetricity condition
\begin{equation}
    \langle\varphi_{\tau}(y)|\hat{v}_1\varphi_{\tau}(y)\rangle = \langle\hat{v}_1
    \varphi_{\tau}(y)|\varphi_{\tau}(y)\rangle
\end{equation}
leads to
\begin{equation}\label{bievol2}
    \overline{\varphi}_{\tau}(\frac{\pi}{b})\varphi_{\tau}(\frac{\pi}{b})-
    \overline{\varphi}_{\tau}(0)\varphi_{\tau}(0)=0.
\end{equation}
We use the result (\ref{bievol}) to calculate the limits
\begin{equation}
    \lim_{y\longrightarrow 0}\varphi_{\tau}(y)=e^{\frac{bk\hbar\tau}{2}}
    \varphi_{0}(0),~~~~
    \lim_{y\longrightarrow \frac{\pi}{b}}\varphi_{\tau}(y)=e^{-\frac{bk\hbar\tau}
    {2}}\varphi_{0}(\frac{\pi}{b}),
\end{equation}
which turns (\ref{bievol2}) into
\begin{equation}\label{biany}
    \overline{\varphi}_0(\frac{\pi}{b})\varphi_0(\frac{\pi}{b})e^{-bk\hbar\tau}-
    \overline{\varphi}_0(0)\varphi_0(0)e^{bk\hbar\tau}=0.
\end{equation}
It is clear that (\ref{biany}) can be satisfied   $\,\forall \tau$
iff $\,{\varphi}_0(\frac{\pi}{b})=0=\varphi_0(0)$. States
with such a property belong to the domain of $\breve{w}$  defined
by (\ref{bispec}).

In order to construct the ``evolving states'' that vanish at the
boundaries, consider the basis vectors $f_n(y)=e^{i2bny}$. Then,
$f_n(y)-f_m(y)$ satisfy the condition (\ref{biany}). Making use of
(\ref{bievol}) we get
\begin{equation}
    f_n(y,\tau)=\bigg(\frac{i-e^{bk\hbar\tau}\tan(\frac{by}{2})}
    {i+e^{bk\hbar\tau}
    \tan(\frac{by}{2})}\bigg)^{2n}
    \sqrt{\frac{1+\tan^2(\frac{by}{2})}{e^{-bk\hbar\tau}+
    e^{bk\hbar\tau}\tan^2(\frac{by}{2})}},
\end{equation}
where $f_n(y,\tau):= f_{n,\tau}(y)$. Moreover we have
\begin{eqnarray}\nonumber
    -ia\frac{d}{dy}f_n(y,\tau)= -i\frac{ab}{2}\big(1+\tan^2(\frac{by}{2})\big)f_n(y,\tau)
    \frac{1}{1+e^{2bk\hbar\tau}\tan^2(\frac{by}{2})}\times \\
    \times\bigg(\frac{(1-e^{2bk\hbar\tau})\tan(\frac{by}{2})}{1+\tan^2(\frac{by}{2})}+
    i4ne^{bk\hbar\tau}\bigg).
\end{eqnarray}
Using the substitution $x=\tan(\frac{by}{2})$ we get
\begin{eqnarray}\nonumber
    \langle f_m|-ia\frac{d}{dy}f_n\rangle=\\
    -ia\int_0^{\infty}
    \bigg(\frac{i-e^{bk\hbar\tau}x}{i+e^{bk\hbar\tau}x}\bigg)^{2(n-m)}
    \frac{(e^{-bk\hbar\tau}-e^{bk\hbar\tau})x}{(e^{-bk\hbar\tau}+e^{bk\hbar\tau}x^2)^2}
    ~dx\\ \nonumber
    +4an\int_0^{\infty}
    \bigg(\frac{i-e^{bk\hbar\tau}x}{i+e^{bk\hbar\tau}x}\bigg)^{2(n-m)}
    \frac{1+x^2}{(e^{-bk\hbar\tau}+e^{bk\hbar\tau}x^2)^2}~dx .
\end{eqnarray}
Another substitution $z=e^{bk\hbar\tau}x$ leads to
\begin{eqnarray}\nonumber
    \langle f_m|-ia\frac{d}{dy}f_n\rangle=\\
    -ia(e^{-bk\hbar\tau}-e^{bk\hbar\tau})\int_0^{\infty}
    \bigg(\frac{i-z}{i+z}\bigg)^{2(n-m)}\frac{z}{(1+z^2)^2}
    ~dz\\ \nonumber
    +4an\int_0^{\infty}
    \bigg(\frac{i-z}{i+z}\bigg)^{2(n-m)}\frac{e^{bk\hbar\tau}+
    e^{-bk\hbar\tau}z^2}{(1+z^2)^2}
    ~dz
\end{eqnarray}
Finally, we obtain
\begin{equation}
 \langle f_m|-ia\frac{d}{dy}f_n\rangle= \left\{
\begin{array}{l}
  \frac{ia}{4(n-m)^2-1}(1-8n(n-m))\sinh(bk\hbar\tau),~~n\neq m \\
  ia\sinh(bk\hbar\tau)+2\pi na\cosh(bk\hbar\tau),~~n= m .\\
\end{array}\right.
\end{equation}
Now, let us introduce
$g_{nm}(y,\tau):=\frac{f_n(y,\tau)-f_m(y,\tau)}{\sqrt{\frac{2\pi}{b}}}$
so that $\|g_{nm}\|=1$. One has
\begin{equation}\label{biexpec}
 \langle g_{nm}|-ia\frac{d}{dy}g_{nm}\rangle=(n+m)ab\cosh(bk\hbar\tau)=\frac{n+m}{2}
 \Delta\cosh(bk\hbar\tau).
\end{equation}

The expectation value of the operator (\ref{biexpec}), defining the
volume operator, is similar to the classical form (\ref{bivi5}). The
vectors $g_{nm}$ may be used in the construction of a basis of the
space of states such that ${\varphi}_0(\frac{\pi}{b})= 0
=\varphi_0(0)$.

\subsection{Summary}

As in the FRW case, resolution of the singularity in the Bianchi I model is due to the loop modification of the Hamiltonian at the classical level. This modification is parameterized by a free continuous parameter $\lambda$.

The spectrum of the volume operator, parameterized by $\lambda$,
is bounded from below and discrete.  An evolution of the
expectation value of the volume operator is similar to the
classical case. We have presented the evolution of only a single
directional volume operator. One may try to generalize this
procedure to the total volume operator. In the case of the Kasner-like
analyzes of dynamics are complicated.

We introduced the so-called true Hamiltonian which proves an independence of the spectrum of the volume operator on the evolution.

\def\baselinestretch{1}

\nonumchapter{Conclusions}

\def\baselinestretch{1.66}


Firstly, in the nonstandard LQC the results are obtained directly on the physical Hilbert space. Secondly, our nonstandard loop quantum cosmology, successfully applied so far to the FRW and Bianchi I models, seems to be highly efficient. For example, only analytical calculations are needed to obtain the results.

Turning the Big Bang into the Big Bounce in our method is due to the modification of the Hamiltonian at the classical level by making use of the loop geometry. The modification is parameterized by a continuous parameter $\lambda$, which value is not known.

In both considered models the spectrum of the volume operators, parameterized by $\lambda$,
are bounded from below and discrete. An evolution of the
expectation values of the volume operators are similar to the
classical cases.

In the Bianchi I model, the phase space of the system is divided into the two
distinct regions: the Kasner-like and the Kasner-unlike. Domains, spectra and eigenvectors of self-adjoint directional volumes, and total volume operators were identified in the
Kasner-unlike case. The peculiarity of the Kasner-like case was identified due to complicated boundary of the phase space region. We propose to overcome this problem by dividing this region further into smaller regions, but with simpler boundaries.
Given a small subregion for the Kasner-like case, we propose a
canonical redefinition of phase space coordinates in such a
way, that we can arrive at relatively simple form of volume
operator and at the same time can simply encode the boundary of
the region into the Schr\"odinger representation. Then, from a
number of different operator orderings the simplest one was chosen.
Domain, spectrum and eigenvectors of the volume operator was founded.
The spectrum is given in an implicit form in terms of special
functions.

Discreteness of space at the quantum level may lead to a
foamy structure of spacetime at the semi-classical level. The discreteness is also specific to the FRW case \cite{Malkiewicz:2009qv}. The difference is that in the Bianchi I
case the variety of possible quanta of a volume is much richer. On
the other hand, the Bianchi type cosmology seems to be more
realistic than the FRW case, near the cosmological singularity.
Thus, an expected foamy structure of space may better fit
cosmological data. Various forms of discreteness of space may
underly many approaches in fundamental physics. So its examination
may be valuable.

As we know $\lambda$ is a free parameter. Without specific choice of $\lambda$, the Big Bounce may occur at any low or high density. The former case (big $\lambda$) contradicts the data of observational cosmology (there was no Big Bounce in the near past) and leads to weakly controlled modification of the expression for the curvature $F^k_{ab}$, i.e. gravitational part of the Hamiltonian (see Appendix A). On the other hand the latter case (small $\lambda$) gives much better approximation for the classical Hamiltonian (see Appendix A), but may easily lead to densities much higher than the Planck scale
density, where the classical formalism is believed to be inadequate. Finding specific value of the parameter $\lambda$, i.e. the energy scale specific to the Big Bounce is an open problem.
It may happen, that the value of the parameter $\lambda$ cannot be  determined, for some
reason, theoretically. The story may turn out to be similar to the case of the short-range
repulsive part of the potential of the nucleon-nucleon interaction introduced to explain the scattering data \cite{RJ} and the nuclear matter saturation of energy \cite{BM}. In such a
case $\lambda$ will become a phenomenological variable parameterizing our ignorance of microscopic properties  of the Universe. Fortunately, there is a rapidly growing
number of data coming from observational cosmology that may be useful in this context.
The cosmic projects for the detection of gamma ray bursts may reveal that the velocity
of cosmic photons depend on their wave lengths, which may be ascribed to the foamy nature of spacetime \cite{RB,Albert:2007qk,Lamon:2008es}. Such dependence is weak, but may sum up to give a measurable effect in the case of photons travelling over cosmological distances across the Universe \cite{GAC}. Presently, available data suggest that such dispersion effects do not occur up to the energy scale $5\times 10^{17}$ GeV \cite{Aharonian:2008kz} so such effects may be present, but at higher energies. Another way to determine the
phenomenological value of the parameter $\lambda$ is the detection of the primordial gravitational waves created at the Big Bounce \cite{Mielczarek:2008pf,Calcagni:2008ig,Grain:2009kw,Grain:2009cj}.

In our method an evolution parameter $\varphi$ does not belong to the physical phase space, contrary to the standard LQC. Thus, it stays classical during the quantization process as well. At quantum level of the Bianchi I model, for the first time in our method, the so-called true Hamiltonian was introduced. It generates a flow in the family of volume quantities, enumerated by an evolution parameter. Having the true Hamiltonian, we could introduce an unitary operator with the evolution parameter $\tau \in \dR$.

It is clear that the next step, in the road to understand an early Universe, is the nonstandard LQC quantization of the Bianchi II cosmological model.

\smallskip


\appendix

\def\baselinestretch{1}

\chapter{Holonomy corrections}

The curvature  of  $SU(2)$ connection $F^k_{ab} =
\partial_a A^k_b - \partial_b A^k_a + \epsilon^k_{ij} A^i_a
A^j_b$, entering the expression (\ref{mothamG}) for the gravitational
part of the Hamiltonian, can be expressed in terms of holonomies.
Using the mean-value and Stokes' theorems we
have\begin{equation}\label{stokes}
    \tau_k\,F^k_{ab}(\vec{x})\approx\frac{1}{s^\sigma_{ab}}\int_\sigma\,
    \tau_k\,F^k_{cd}\,d x^c\wedge
    dx^d\approx\frac{1}{s^\sigma_{ab}}\Big( \mathcal{P}\exp\big(\oint_{\partial \sigma}
    \tau_k\,A^k_c\,dx^c\big) - 1\Big),
\end{equation}
where $\partial \sigma$ is the boundary of a small surface
$\sigma$ with center at $\vec{x}$, and where $s^\sigma_{ab}:=
\int_\sigma dx^a\wedge dx^b$. The expression for $F^k_{ab}$ is
exact  but in the limit when we shrink the area enclosed by the
loop $\partial \sigma$ to zero. If we choose $\partial \sigma$ in
the form of the square  $\square_{ij}$ with sides length $\lambda$,
the expression for a small value of $\lambda = \mu_0$ has the
form \cite{Mielczarek:2008zz}
\begin{equation}\label{finite}
F^k_{ab}(\mu_0)= \lim_{\lambda\,\rightarrow \,\mu_0}
\Big\{-2\;Tr\;\Big(\frac{h^{(\lambda)}_{\Box_{ij}}-1}{\lambda^2
V_o^{2/3}}\Big)\;{\tau^k}\; ^o\omega^i_a  \; ^o\omega^j_a +
\frac{\mathcal{O}(\lambda^4)}{\lambda^2}\Big\},
\end{equation}
and we have
\begin{equation}\label{zero}
F^k_{ab}= \lim_{\mu_0\,\rightarrow \,0}\, F^k_{ab}(\mu_0).
\end{equation}

In the standard LQC the $\mathcal{O}(\lambda^4)$ holonomy corrections
are ignored (see, e.g. \cite{Ashtekar:2006wn,Ashtekar:2007em}). It
was found in \cite{Mielczarek:2008zz,Hrycyna:2008yu} that
including higher order corrections leads to new curvature
singularities different from the initial singularity and increases
an ambiguity problem of loop cosmology. However, the holonomy
corrections do not change  the result that the Big Bounce is a
consequence of the loopy nature of geometry \cite{Chiou:2009hk}.

Taking only the first term of (\ref{finite}) leads to the simplest
modification of gravity, but may be insufficient for the
description of the inflationary phase. The choice of $\mu_0$ based
on the expectation that the Big Bounce should occur at the Planck
scale \cite{Ashtekar:2006wn} has little justification
\cite{Bojowald:2008ik}. The significance of Planck's scale for
quantum gravity seems to be rather a belief than proved result
(see, e.g. \cite{Meschini:2006gm}). Heuristic reasoning playing
game at the same time with Heisenberg's uncertainty principle,
Schwarzschild's radius and process of measurement cannot replace a
proof  (see, e.g. \cite{WP}).

\chapter{Justification for using observables to classify phase space}

\section{Nonregularized case}

The nonregularized case means that we do not modify general relativity by loop geometry
(loops of finite length). For this case one has\footnote{obtained from (\ref{biO2}) in the limit $\lambda \rightarrow 0$}
\begin{equation}\label{non1}
\textrm{O}_i= \frac{1}{\kappa\gamma}v_i\beta_i ,
\end{equation}
where
\begin{equation}\label{non2}
v_i:= (a_ja_kL_jL_k)^{3/2},~~~~~~\beta_i:=
\frac{\dot{a}_i\gamma}{(a_ja_kL_jL_k)^{1/2}}.
\end{equation}
It is clear that
\begin{equation}\label{non3}
\textrm{O}_i=
\frac{1}{\kappa}a_ia_ja_kL_iL_jL_k\frac{\dot{a}_i}{a_i}=:
\frac{V}{\kappa}H_i,
\end{equation}
which leads to
\begin{equation}\label{non4}
H_k= \frac{\kappa}{V}\textrm{O}_k
\end{equation}
where $H_k:= \dot{a}_k/a_k$ is a directional Hubble parameter in the $k$-th direction. In what follows we use $H_k$ to describe contraction or expansion of spacetime.
It results from (\ref{non4}) that classical dynamics of our system in the nonregularized case can be described in terms of the observables $\textrm{O}_k$.

\section{Regularized case}

In this case we modify gravity by loop geometry, due to
(\ref{bivi2}), we have
\begin{equation}\label{e6}
v_i= \kappa\gamma\lambda|\textrm{O}_i|\,\cosh\big(
3\kappa\big((\textrm{O}_j+\textrm{O}_k )\,\varphi -
\textrm{A}_i\big)\big)
\end{equation}
so we get
\begin{equation}\label{e1}
\frac{\dot{v_i}}{v_i}= 3\kappa\,(\textrm{O}_j +
\textrm{O}_k)\,\tanh\big( 3\kappa\big((\textrm{O}_j+\textrm{O}_k
)\,\varphi - \textrm{A}_i\big)\big)\,\dot{\varphi}
\end{equation}
where $\dot{\varphi}= \frac{1}{2}\,\frac{1}{V}$.
On the other hand we have
\begin{equation}\label{e2}
\frac{\dot{v_i}}{v_i}= \frac{3}{2}\,(H_j + H_k).
\end{equation}

Comparing (\ref{e1}) and (\ref{e2}) we obtain
\begin{equation}\label{e3}
H_j + H_k = \frac{\kappa}{V}\,(\textrm{O}_j + \textrm{O}_k)\,\tanh\big(
3\kappa\big((\textrm{O}_j+\textrm{O}_k )\,\varphi -
\textrm{A}_i\big)\big).
\end{equation}
Denoting the moment of occurring a bounce by $\varphi_i^{\text{B}}$,
defined by
\begin{equation}\label{e5}
\textrm{A}_i= (\textrm{O}_j + \textrm{O}_k)\,\varphi_i^{\text{B}},
\end{equation}
we obtain the formula
\begin{equation}\label{e4}
H_j + H_k= \frac{\kappa}{V}\,(\textrm{O}_j + \textrm{O}_k)\,\tanh\big(
3\kappa\,(\textrm{O}_j+\textrm{O}_k )\,(\varphi -
\varphi_i^{\text{B}})\big).
\end{equation}

For the purpose of interpretation it is convenient to rewrite
(\ref{e4}) in different form. Since $\cosh(x)^2 - \sinh(x)^2= 1$,
we have
\begin{equation}
\tanh(x)= \pm\sqrt{1 - \frac{1}{\cosh^2(x)}}.
\end{equation}
In the case of expanding universe, $\varphi>\varphi_i^{\text{B}}$, we
have the expression
\begin{equation}
\tanh\big( 3\kappa\,(\textrm{O}_j+\textrm{O}_k )\,(\phi -
\phi_i^{\text{B}})\big)= \sqrt{1 - \frac{1}{\cosh^2 \big(
3\kappa\,(\textrm{O}_j+\textrm{O}_k )\,(\varphi -
\varphi_i^{\text{B}})\big)}}.
\end{equation}
It results from (\ref{e6}) and (\ref{e5}) that
\begin{equation}
v_i= \kappa\gamma\lambda\textrm{O}_i\,\cosh\big(
3\kappa\,(\textrm{O}_j+\textrm{O}_k)\,(\varphi -
\varphi_i^{\text{B}})\big).
\end{equation}
Since $v_i^\text{B}= \kappa\gamma\lambda\textrm{O}_i$ (directional
volume at the i-th bounce), we have
\begin{equation}
\frac{1}{\cosh\big(3\kappa\,(\textrm{O}_j+\textrm{O}_k)\,(\varphi -
\varphi_i^{\text{B}})\big)}= \frac{v_i^\text{B}}{v_i},
\end{equation}
which means that
\begin{equation}\label{e7}
\tanh\big( 3\kappa\,(\textrm{O}_j+\textrm{O}_k )\,(\varphi -
\varphi_i^{\text{B}})\big)= \sqrt{1 -
\bigg(\frac{v_i^\text{B}}{v_i}\bigg)^2}.
\end{equation}
Inserting (\ref{e7}) into (\ref{e4}) gives finally
\begin{equation}\label{e8}
H_j + H_k= \frac{\kappa}{V}\,(\textrm{O}_j + \textrm{O}_k)\,\sqrt{1 -
\bigg(\frac{v_i^\text{B}}{v_i}\bigg)^2}.
\end{equation}
It results from (\ref{e8}) that for very large $v_i$, comparing
to $v_i^{B}$, we can apply the approximation
\begin{equation}\label{e9}
H_j + H_k= \frac{\kappa}{V}\,(\textrm{O}_j + \textrm{O}_k),
\end{equation}
which finally leads to
\begin{equation}\label{e10}
H_k = \frac{\kappa}{V}\,\textrm{O}_k.
\end{equation}

Comparing (\ref{non4}) and (\ref{e10}) one can see that also for the regularized case the structure of the classical phase space may be described in terms of $\textrm{O}_k$ observables, but only when we consider the Universe far away from the Big Bounce ($v_i\gg v_i^{B}$). However, for our analyzes that limitation does not matter.

\chapter{Non-uniqness in quantization of the Bianchi I}

It is clear that quantum cosmology calculations are plagued by quantization ambiguities. For example, there exists a huge freedom in ordering of elementary operators defining compound observables, which may lead to different quantum operators. Classical commutativity of variables does not extend to corresponding quantum operators. Other ambiguities are discussed below. Such ambiguities can be largely reduced when some quantum data from cosmological observations become available. Confrontation of theoretical predictions against these data would enable finding realistic quantum cosmology models.

\section{Unitarily non-equivalent volume operators}

In both Kasner-like and Kasner-unlike cases, we have reduced the
Hilbert space by removing the double degeneracy of
eigenvalues for the volume operators (see the discussion after
equations (\ref{biw^2}) and (\ref{biw2})). We have used the ``natural''
condition that the wave function should vanish at the boundaries
of an interval. However, there are also other mathematically
well-defined choices for the boundary conditions. We will
demonstrate this non-uniqueness for the Kasner-unlike case.
Similar reasoning applies to another case.

Let us begin with the equation (\ref{biw^2})
\begin{equation}
-a^2\frac{d^2}{dy^2}f=\nu^2f,~~~~y\in (0,\pi/b),
\end{equation}
which has the solution
\begin{equation}
f_{\nu}=N_1\sin(\frac{\nu}{a}y)+N_2\cos(\frac{\nu}{a}y),~~~~N_1,N_2\in
\mathbb{C},
\end{equation}
for each value of $\nu\in \dR_+$ (~$\nu\mapsto -\nu$ does not
produce any new space of solutions). Our task is the determination
of self-adjointness  of $\breve{w}:=\sqrt{-a^2\frac{d^2}{dy^2}}$
and removing  the double degeneracy of eigenvalues. The
symmetricity condition reads
\begin{equation}\label{bisym}
    \int_{I}\bar{f}f''=\bar{f}f'\bigg|^{\pi/b}_{0}-\bar{f}'f\bigg|^{\pi/b}_0
    +\int_I\bar{f}''f.
\end{equation}

We can set:
\begin{itemize}
    \item  $f(0)=f(\pi/b)=0 \Rightarrow
    f_{\nu}=\sin(\frac{\nu}{a}y),~~\nu=ab,2ab,3ab,\dots$
    \item $f'(0)=f'(\pi/b)=0 \Rightarrow
    f_{\nu}=\cos(\frac{\nu}{a}y),~~\nu=0,ab,2ab,3ab,\dots$
    \item $f(0)=f'(\pi/b)=0 \Rightarrow
    f_{\nu}=\sin(\frac{\nu}{a}y),~~\nu=\frac{1}{2}ab,\frac{3}{2}ab,
    \frac{5}{2}ab,\dots$
\end{itemize}
where $ab = 12\pi\hbar G\gamma\lambda.$ All these choices are
non-equivalent, since they lead to different spectra.

\section{Standard quantization}

Let us change the coordinates of the Kasner-like sector phase
space $(\Omega_1,\Omega_2)$, defined by (\ref{bired1}), into a new
canonical pair as follows
\begin{equation}
X:=\sqrt{2\Omega_1}~~\textrm{and}~~P:=\Omega_2\sqrt{2\Omega_1},
\end{equation}
where
\begin{equation}
(X,P)\in (0,\sqrt{2d_1/b})\times\dR,~~~~\{X,P\}=1.
\end{equation}
In the new variables the volume (\ref{bir3}) reads
\begin{equation}
\frac{1}{4\pi G\gamma\lambda}~v_1=\frac{1}{2}P^2+ \frac{1}{2} X^2.
\end{equation}
Thus, in these  variables  the volume has a form of the
Hamiltonian of the harmonic oscillator in a ``box''
$(0,\sqrt{2d_1/b})$.

In the Schr\"odinger representation, i.e. $\hat{X}:=x$ and
$\hat{P}:=-i\hbar\partial_{x}$, a standard quantization yields
\begin{equation}
    \frac{1}{4\pi G\gamma\lambda}~\hat{v}=-\frac{\hbar^2}{2}\partial^2_{xx}
    + \frac{1}{2}x^2 ,
\end{equation}
which corresponds to the ``nonstandard'' quantization (\ref{binov1})
with the parameters $m=k=1/4$ and $y = \sqrt{2} x$ (with $\hbar
=1$).

Thus, we can see that the prescription defined by (\ref{binov}) and
(\ref{binovum}) includes not only a standard prescription, but many
others. As an illustration only one, corresponding to the well known harmonic oscillator, has been completed.

\setlinespacing{1.44}


\begin{thebibliography}{99}

\bibitem{MTW} C. W. Misner, K. S. Thorne and J. A. Wheeler \textit{Gravitation}
(San Francisco: W. H. Freeman and Company, 1973).

\bibitem{SWH} S. W. Hawking and  G. F. R. Ellis \textit{The large scale structure
 of space-time} (Cambridge, Cambridge University Press, 1975).

\bibitem{JPAK} J. Pleba\'{n}ski and A. Krasi\'{n}ski \textit{An Introduction to
 General Relativity and Cosmology} (Cambridge, Cambridge University
 Press, 2006).

\bibitem{JMMS} J.~M.~M.~Senovilla,
 ``Singularity Theorems and their Consequences'', Gen. Rel. Grav.
 {\bf 30} (1998) 701.

\bibitem{BKL1} V. A.~Belinskii, I. M.~Khalatnikov and E. M.~Lifshitz, Adv. Phys.
{\bf 19}, 525 (1970).

\bibitem{BKL2} V. A.~Belinskii and I. M.~Khalatnikov, Sov. Phys. JETP {\bf 36}, 591
(1973).

\bibitem{BKL}
V. A. Belinskii, I. M. Khalatnikov and E. M. Lifschitz, ``A
general solution of the Einstein equations with a time
singularity,'' Adv. Phys. {\bf 31}, 639 (1982).

\bibitem{PrzegBKL1}
 V.~Belinski,
 ``Cosmological singularity'',
 \texttt{arXiv:0910.0374}.

\bibitem{PrzegBKL2}
 A. Yu.~Kamenshchik,
 ``The problem of singularities and chaos in cosmology'',
 \texttt{arXiv:1006.2725}.

\bibitem{Numeryczna1}
 B.K.~Berger,
 ``Numerical Approaches to Spacetime Singularities'',
 Living Rev. in Relativity (2002-1).

\bibitem{PrzegBKL3}
 A.~Ashtekar, A.~Henderson and D.~Sloan,
 ``Hamiltonian General Relativity and the Belinskii, Khalatnikov, Lifshitz Conjecture'',
 \texttt{arXiv:0811.4160}.

\bibitem{BianchiII} A.~Ashtekar and E.~Wilson-Ewing, ``Loop quantum cosmology of Bianchi II type models'', \texttt{arXiv:0910.1278}.

\bibitem{Ashtekar:2009vc}
  A.~Ashtekar and E.~Wilson-Ewing,
  ``Loop quantum cosmology of Bianchi I models,''
  \texttt{arXiv:0903.3397}.

\bibitem{Ashtekar:2003hd}
  A.~Ashtekar, M.~Bojowald and J.~Lewandowski,
  ``Mathematical structure of loop quantum cosmology'',
  Adv.\ Theor.\ Math.\ Phys.\  {\bf 7} (2003) 233,
  \texttt{arXiv:0304074 [gr-qc]}.

\bibitem{Ashtekar:2004eh}
  A.~Ashtekar and J.~Lewandowski,
  ``Background independent quantum gravity: A status report'',
  Class.\ Quant.\ Grav.\  {\bf 21} (2004) R53,
  \texttt{arXiv:0404018 [gr-qc]}.

\bibitem{TT} T. Thiemann \textit{Modern Canonical Quantum General Relativity}
(Cambridge: Cambridge University Press, 2007).

\bibitem{CR} C. Rovelli \textit{Quantum Gravity}
(Cambridge: CUP, 2004).

\bibitem{Bojowald:2001xe}
  M.~Bojowald,
  ``Absence of singularity in loop quantum cosmology,''
  Phys.\ Rev.\ Lett.\  {\bf 86} (2001) 5227,
  \texttt{arXiv:0102069 [gr-qc]}.

\bibitem{Bojowald:2003md}
  M.~Bojowald,
  ``Homogeneous loop quantum cosmology,''
  Class.\ Quant.\ Grav.\  {\bf 20} (2003) 2595,
  \texttt{arXiv:0303073 [gr-qc]}.

\bibitem{Bojowald:2006da}
  M.~Bojowald,
  ``Loop quantum cosmology'',
  Living Rev.\ Rel.\  {\bf 8} (2005) 11,
  \texttt{arXiv:0601085 [gr-qc]}.

\bibitem{Ashtekar:2006uz}
  A.~Ashtekar, T.~Paw{\l}owski and P.~Singh,
  ``Quantum nature of the big bang: An analytical and numerical
  investigation'',
  Phys.\ Rev.\  D {\bf 73} (2006) 124038,
  \texttt{arXiv:0604013 [gr-qc]}.

\bibitem{Ashtekar:2006wn}
  A.~Ashtekar, T.~Paw{\l}owski and P.~Singh,
  ``Quantum nature of the big bang: Improved dynamics'',
  Phys.\ Rev.\  D {\bf 74} (2006) 084003,
  \texttt{arXiv:0607039 [gr-qc]}.

\bibitem{Ashtekar:2007em}
  A.~Ashtekar, A.~Corichi and P.~Singh,
  ``On the robustness of key features of loop quantum cosmology'',
  Phys.\ Rev.\  D {\bf 77} (2008) 024046,
  \texttt{arXiv:0710.3565}.

\bibitem{Przeg1}
 A.~Ashtekar, ``Loop Quantum Cosmology: An Overview'',
 \texttt{arXiv:0812.0177}.

\bibitem{Przeg2}
 A.~Ashtekar,
 ``Singularity Resolution in Loop Quantum Cosmology: A Brief Overview'',
 \texttt{arXiv:0812.4703}.

\bibitem{Przeg3}
 A.~Ashtekar,
 ``The Big Bang and the Quantum'',
  \texttt{arXiv:1005.5491}.

\bibitem{Ashtekar:2008vv}
  A.~Ashtekar,
  ``Quantum Space-times: Beyond the Continuum of Minkowski and
  Einstein'',
  \texttt{arXiv:0810.0514}.

\bibitem{Date:2004fj}
  G.~Date and G.~M.~Hossain,
  ``Genericity of big bounce in isotropic loop quantum cosmology,''
  Phys.\ Rev.\ Lett.\  {\bf 94} (2005) 011302,
  \texttt{arXiv:0407074 [gr-qc]}.

\bibitem{Dzierzak:2008dy}
  P.~Dzierzak, J.~Jezierski, P.~Malkiewicz and W.~Piechocki,
  ``The minimum length problem of loop quantum cosmology'', Acta Phys. \textbf{41} (2010) 717,  \texttt{arXiv:0810.3172}.

\bibitem{PAM} P. A. M. Dirac, \textit{Lectures on Quantum Mechanics}
(New York: Belfer Graduate School of Science Monographs Series,
1964).

\bibitem{HT} M. Henneaux and C. Teitelboim, \textit{Quantization of Gauge Systems}
(Princeton: Princeton University Press, 1992).

\bibitem{Ashtekar:1996eg}
  A.~Ashtekar and J.~Lewandowski,
  ``Quantum theory of geometry. I: Area operators'',
  Class.\ Quant.\ Grav.\  {\bf 14} (1997) A55,
  \texttt{arXiv:9602046 [gr-qc]}.

\bibitem{Rovelli:1994ge}
  C.~Rovelli and L.~Smolin,
  ``Discreteness of area and volume in quantum gravity'',
  Nucl.\ Phys.\  B {\bf 442} (1995) 593
  [Erratum-ibid.\  B {\bf 456} (1995) 753],
  \texttt{arXiv:9411005 [gr-qc]}.

\bibitem{Marolf:2000iq}
  D.~Marolf,
  ``Group averaging and refined algebraic quantization: Where are we
  now?'',
  \texttt{arXiv:0011112 [gr-qc]}.

\bibitem{Ashtekar:1995zh}
  A.~Ashtekar, J.~Lewandowski, D.~Marolf, J.~Mourao and T.~Thiemann,
  ``Quantization of diffeomorphism invariant theories of connections with local
  degrees of freedom'',
  J.\ Math.\ Phys.\  {\bf 36} (1995) 6456,
  \texttt{arXiv:9504018 [gr-qc]}.

\bibitem{Bojowald:2008ik}
  M.~Bojowald,
  ``Consistent Loop Quantum Cosmology'',
  Class.\ Quant.\ Grav.\  {\bf 26} (2009) 075020,
  \texttt{arXiv:0811.4129}.

\bibitem{Dzierzak:2009ip}
  P.~Dzierzak, P.~Malkiewicz and W.~Piechocki,
  ``Turning Big Bang into Big Bounce: I. Classical Dynamics,''
  Phys.\ Rev.\  D {\bf 80}, 104001 (2009),
  \texttt{arXiv:0907.3436}.

\bibitem{Malkiewicz:2009qv}
  P.~Malkiewicz and W.~Piechocki,
  ``Turning big bang into big bounce: II. Quantum dynamics,''
  \texttt{arXiv:0908.4029}.

 \bibitem{Mielczarek:2010rq}
    J.~Mielczarek and W.~Piechocki,
    ``Observables for FRW model with cosmological constant in the framework of
    loop cosmology,''
    Phys.\ Rev.\  D {\bf 82} (2010) 043529,
    \texttt{arXiv:1001.3999}.

\bibitem{Giesel:2007wn}
  K.~Giesel and T.~Thiemann,
  ``Algebraic Quantum Gravity (AQG) IV. Reduced Phase Space Quantisation of
  Loop Quantum Gravity,''
  Class.\ Quant.\ Grav.\  {\bf 27} (2010) 175009,
  \texttt{arXiv:0711.0119}.

\bibitem{Chiou:2007mg}
  D.~W.~Chiou,
  ``Effective Dynamics, Big Bounces and Scaling Symmetry in Bianchi Type I Loop
  Quantum Cosmology,''
  Phys.\ Rev.\  D {\bf 76} (2007) 124037,
  \texttt{arXiv:0710.0416}.

\bibitem{Malkiewicz:2009zd}
  P.~Malkiewicz and W.~Piechocki,
  ``Energy Scale of the Big Bounce,''
   Phys.\ Rev.\  D {\bf 80}, 063506 (2009),
  \texttt{arXiv:0903.4352}.

\bibitem{RS} M. Reed and B. Simon, {\it Methods of Modern Mathematical
Physics} (Academic Press, San Diego, 1975).

\bibitem{DS} N. Dunford and J. T. Schwartz, {\it Linear Operators}
(Interscience Publishers, New York, 1958).

\bibitem{Khalatnikov:2008zt}
  I.~M.~Khalatnikov and A.~Y.~Kamenshchik,
  ``Lev Landau and the problem of singularities in cosmology,''
  \texttt{arXiv:0803.2684}.

\bibitem{Montani:2007vu}
  G.~Montani, M.~V.~Battisti, R.~Benini and G.~Imponente,
  ``Classical and Quantum Features of the Mixmaster Singularity,''
  Int.\ J.\ Mod.\ Phys.\  A {\bf 23} (2008) 2353,
  \texttt{arXiv:0712.3008}.

\bibitem{Garfinkle:2007rv}
  D.~Garfinkle,
  ``Numerical simulations of general gravitational singularities,''
  Class.\ Quant.\ Grav.\  {\bf 24} (2007) S295,
  \texttt{arXiv:0808.0160}.

\bibitem{Erickson:2003zm}
  J.~K.~Erickson, D.~H.~Wesley, P.~J.~Steinhardt and N.~Turok,
  ``Kasner and mixmaster behavior in universes with equation of state w $\geqslant$
  1,''  Phys.\ Rev.\  D {\bf 69} (2004) 063514,
  \texttt{arXiv:0312009 [hep-th]}.

\bibitem{Dzierzak:2009dj}
  P.~Dzierzak and W.~Piechocki,
  ``Bianchi I model in terms of non-standard LQC: Classical dynamics,''
  Phys.\ Rev.\  D {\bf 80}, 124033 (2009),
  \texttt{arXiv:0909.4211}.

\bibitem{all}
P.~Malkiewicz, W.~Piechocki and P.~Dzierzak
  ``Bianchi I model in terms of non-standard loop quantum cosmology: II. Quantum dynamics'',  submitted for publication, \texttt{arXiv:1010.2930}.

\bibitem{Chiou:2006qq}
  D.~W.~Chiou,
  ``Loop Quantum Cosmology in Bianchi Type I Models: Analytical
  Investigation,''
  Phys.\ Rev.\  D {\bf 75} (2007) 024029,
  \texttt{arXiv:0609029 [gr-qc]}.

\bibitem{Szulc:2008ar}
  L.~Szulc,
  ``Loop quantum cosmology of diagonal Bianchi type I model: Simplifications
  and scaling problems,''
  Phys.\ Rev.\  D {\bf 78} (2008) 064035,
  \texttt{arXiv:0803.3559}.

\bibitem{MartinBenito:2008wx}
  M.~Martin-Benito, G.~A.~Mena Marugan and T.~Pawlowski,
  ``Loop Quantization of Vacuum Bianchi I Cosmology,''
  Phys.\ Rev.\  D {\bf 78} (2008) 064008,
  \texttt{arXiv:0804.3157}.

\bibitem{RJ} R.~Jastrow, ``On the Nucleon-Nucleon Interaction'',
 Phys.\ Rev. {\bf 81}  (1951) 165.

\bibitem{BM} A. A. Bohr and B. R. Mottelson \textit{Nuclear Structure}
 Vol. 1 (New York: Benjamin, Inc, 1969).

\bibitem{RB} F. Aharonian {\it et al.}, ``Limits on an energy
dependence of the speed of light from a flare of the active galaxy
PKS 2155-304'', Phys.\ Rev.\ Lett. {\bf 101} (2008) 170402.

\bibitem{Albert:2007qk}
  J.~Albert {\it et al.}
  ``Probing quantum gravity using photons from a flare of the active galactic
  nucleus Markarian 501 observed by the MAGIC telescope''
  Phys.\ Lett.\  B {\bf 668} (2008) 253,
  \texttt{arXiv:0708.2889}.

\bibitem{Lamon:2008es}
  R.~Lamon,  ``GLAST and Lorentz violation'',  JCAP {\bf 0808} (2008) 022,
  \texttt{arXiv:0805.1219}.

\bibitem{GAC}
 G.~Amelino-Camelia {\it et al.}, ``Tests of quantum gravity from observations
 of bold gamma-ray bursts'', Nature {\bf 393}, 763 (1998).

\bibitem{Aharonian:2008kz}
  F.~Aharonian {\it et al.},
  ``Limits on an Energy Dependence of the Speed of Light from a Flare of the
  Active Galaxy PKS 2155-304'',
  Phys.\ Rev.\ Lett.\  {\bf 101}, 170402 (2008),
  \texttt{arXiv:0810.3475}.

\bibitem{Mielczarek:2008pf}
  J.~Mielczarek,
  ``Gravitational waves from the Big Bounce'',
  JCAP {\bf 0811} (2008) 011,
  \texttt{arXiv:0807.0712}.

\bibitem{Calcagni:2008ig}
  G.~Calcagni and G.~M.~Hossain,
  ``Loop quantum cosmology and tensor perturbations in the early
  universe'',
  \texttt{arXiv:0810.4330}.

\bibitem{Grain:2009kw}
  J.~Grain and A.~Barrau,
  ``Cosmological footprints of loop quantum gravity,''
  Phys.\ Rev.\ Lett.\  {\bf 102} (2009) 081301,
  \texttt{arXiv:0902.0145}.

\bibitem{Grain:2009cj}
  J.~Grain, A.~Barrau and A.~Gorecki,
  ``Inverse volume corrections from loop quantum gravity and the primordial
  tensor power spectrum in slow-roll inflation,''
  Phys.\ Rev.\  D {\bf 79} (2009) 084015,
  \texttt{arXiv:0902.3605}.

\bibitem{Mielczarek:2008zz}
  J.~Mielczarek and M.~Szydlowski,
  ``Emerging singularities in the bouncing loop cosmology'',
  Phys.\ Rev.\  D {\bf 77} (2008) 124008,
  \texttt{arXiv:0801.1073}.

\bibitem{Hrycyna:2008yu}
  O.~Hrycyna, J.~Mielczarek and M.~Szydlowski,
  ``Effects of the quantisation ambiguities on the Big Bounce
  dynamics'',
  \texttt{arXiv:0804.2778}.

\bibitem{Chiou:2009hk}
  D.~W.~Chiou and L.~F.~Li,
  ``How loopy is the quantum bounce? A heuristic analysis of higher order
  holonomy corrections in LQC'',
  \texttt{arXiv:0901.1757}.

\bibitem{Meschini:2006gm}
 D.~Meschini,
``Planck-scale physics: Facts and beliefs'',
 Found.\ Sci.\  {\bf 12} (2007) 277,
 \texttt{arXiv:0601097 [gr-qc]}.

\bibitem{WP}
W.~Piechocki, ``Structure of spacetime at the Planck scale'', Acta
Phys. Polon. B {\bf 21} (1990) 711.




























\end{thebibliography}
\end{document}